\RequirePackage{lineno}
\documentclass[%
 preprint,
superscriptaddress,
 amsmath,amssymb,
 aps,
 prf,
]{revtex4-2}

\usepackage{tikz}
\usetikzlibrary{plotmarks}
\newcommand\marksymbol[2]{\tikz[#2,scale=1.2]\pgfuseplotmark{#1};}
\usepackage{ cuted}
\usepackage{color}
\usepackage{graphicx}% Include figure files
\usepackage{dcolumn}% Align table columns on decimal point

\usepackage{bm}
\usepackage{subfig}% bold math
\usepackage{multirow,stackengine}
\usepackage{braket}
\graphicspath{ {./Figure/} }
\usepackage{color}

\newcommand{\red}[1]{{\color{black}#1}}
\usepackage[hidelinks,colorlinks=true,linkcolor=blue,citecolor=blue]{hyperref}
\newcommand{\redsec}[1]{{\color{black}#1}}
\begin{document}
\modulolinenumbers[5]
% \linenumbers
% \prep
% \Large
\title{Composite active drag control   in turbulent channel flows}

\author{Jie Yao}
\email[Email:]{jie.yao@ttu.edu}
\affiliation{
Texas Tech University, Department of Mechanical Engineering, Lubbock, Texas, USA, 79409
}

\author{Xi Chen}
 \affiliation{
Key Laboratory of Fluid Mechanics of Ministry of Education, Beihang University, Beijing, People’s Republic of China, 100191 }

\author{Fazle Hussain}
\affiliation{
Texas Tech University, Department of Mechanical Engineering, Lubbock, Texas, USA, 79409
}
\date{\today}% It is always \today, today,

\begin{abstract}
A composite drag control (CDC) combining the opposition (OC) and spanwise opposed wall-jet forcing (SOJF)
methods is studied  in a turbulent channel flow via direct numerical simulation of the incompressible Navier–Stokes equations.
A maximum drag reduction of about $33\%$ is obtained for CDC -- much higher than that produced by either individual method (namely, 19\% for SOJF and 23\% for OC).
Due to the small power input required for both OC and SOJF methods, a significant net power saving (about 32\%) is achieved via CDC.
% Flow analysis reveals that the composite control can take advantage of both  OC and SOJF . 
Flow analysis shows that CDC can take advantage of both OC and SOJF methods to better suppress drag producing, near-wall turbulent structures -- vortices and streaks. In particular,  due to the presence of the large-scale coherent swirls generated by SOJF, it is more effective than OC in suppressing the  random turbulence. Moreover, due to the  OC's role in  suppressing random small-scale turbulence, CDC requires  weaker large-scale coherent swirls than those  using SOJF only -- hence decreasing the drag contribution associated with large-scale swirls. 
In summary, our results suggest  prospects  of employing composite control strategy for effective skin friction drag reduction, particularly at very high Reynolds numbers.

\end{abstract}

\maketitle

% \linenumbers

\section{Introduction}

As the skin friction drag of aircraft and watercraft constitutes  a  large  fraction  of  the  total  aerodynamic  drag (e.g., approximately 50\% of  aircraft and about $90\%$ of ships and submarines),  its reduction   is of great significance for  energy saving and  environmental protection.
Benefited from the continuing improved understanding of flow dynamics \cite{waleffe1997self,jimenez1999autonomous, schoppa2002coherent}, especially the role of near-wall organized coherent structures, various successful drag control strategies have been proposed. 
In general, depending on whether external energy is required to drive the control,  these strategies fall into two categories \cite{gad2007flow}: passive and active.
For the passive method,  riblets are one of the highly investigated approaches and  have been shown to yield about $5-7\%$ drag reduction in the full-size trials \citep{bechert1997experiments}.
However, due to the specific requirement, sustenance of the tiny riblets against erosion by atmospheric dust in practical applications is still a challenge.
Recently, drag control using superhydrophobic surfaces  has been conceived and developed. It  has been reported  to yield up to 50\% and 75\% drag reduction in laboratory turbulent channel  \citep{rothstein2010slip}
and boundary layer \citep{park2014superhydrophobic} experiments, respectively.
However, as the superhydrophobic surfaces are severely vulnerable to high pressure and high shear rate  \citep{checco2014collapse,rastegari2018common},  they  have not yet evolved into a practical means.

As for active control methods, they can be further grouped into open- and closed-loop techniques, based on whether sensing is required or not. 
As one of the simplest open-loop techniques,  uniform blowing and suction has been extensively investigated, where uniform blowing is found to reduce the skin friction drag, while uniform suction  increases it \cite{sumitani1995direct,kametani2011direct}. 
The streamwise traveling wave-based (also including spanwise wall oscillations) control strategies are fascinating \citep{jung1992suppression,choi2002drag,quadrio2009streamwise,Quadrio2011, Agostini2014,yakeno2014modification}, where significant drag reductions  are achieved \citep{choi1994active,kim2007linear,chung2011effectiveness,deng2012, mamori2014effect}. 
However, considering the typically large amount of power   required for actuation, whether effective drag reduction can be obtained at a reasonably high Reynolds number ($Re$) remains unclear \cite{gatti2016reynolds,yao2019reynolds}. 
Streamwise  traveling  wave-like  wall  blowing  and  suction is another  typical  example  of  open-loop  control  techniques. With proper selection of  the traveling wave parameters, significant  drag  reduction  and  even  sub-laminar drag reduction  were found  in  turbulent channel flows \cite{min2006sustained,moarref2010controlling,mamori2014effect,kaithakkal2020dissimilarity}. 

One of the promising  closed-loop  control techniques is the  opposition control (OC), which was first studied by \citeauthor{choi1994active} \cite{choi1994active}. 
Based on the measurements in a plane  parallel to the wall (called ``detection plane''), OC employs local wall blowing and suction to counteract the sweep and ejection motions induced by the energetic near-wall  streamwise vortices. Approximately 25\% drag reduction was achieved in turbulent channel flow at friction Reynolds number $Re_\tau(\equiv u_\tau h/\nu)\approx110$  with the detection plane located at $y^+_d=10$ \cite{choi1994active}. [$u_\tau$ is the friction velocity, $\nu$ is the kinematic viscosity, and $h$ is the half channel height; the superscript $+$
indicates that the quantity is scaled by the viscous wall units, namely, $y^+=yu_\tau /\nu$.] 
\citeauthor{hammond1998observed} \cite{hammond1998observed} later  found that the maximum drag reduction  is obtained when  $y^+_d=15$.
For a practical  application, sensors and actuators need to  be  wall-based  and  flush-mounted  to  avoid  parasitic  drag.
\citeauthor{lee1998suboptimal} \cite{lee1998suboptimal} developed the suboptimal control that employs information measurable at the wall, e.g., using streamwise wall shear stress, spanwise wall shear stress, and wall pressure.
As machine learning is becoming  one of today’s most rapidly growing technique, \citeauthor{han2020active}
\cite{han2020active} and \citeauthor{park2020machine} \cite{park2020machine}  recently pursued the feasibility of employing the convolutional neural network (CNN) to predict the wall-normal velocities on the detection plane using either spanwise or streamwise wall shear stress obtained from the direct numerical simulation of channel flow.
Applying the trained CNNs to  turbulent channel flows at low $Re$, significant amounts of drag reduction   can be  achieved.
In addition, \citeauthor{park2020machine} \cite{park2020machine} also found that $15\%$ drag reduction can still be obtained at $Re_\tau=578$ by applying the CNN trained at $Re_\tau=180$. 

Currently, most of the above-mentioned drag control strategies are targeted at  interrupting   the  self-sustaining cycle \cite{waleffe1997self,schoppa2002coherent}, where 
the low-speed streaks are generated via the lift-up effect of faster advecting streamwise vortices leaving streaks as their trails, while the instability/transient growth of low-speed streaks – if sufficiently strong –  regenerates the streamwise vortices.
At  practical, high $Re$'s, these near-wall features are physically very small; hence  effective control of them becomes rather challenging  due to the tiny size of the appropriate sensors and actuators.
In addition, as $Re$ increases, the role of outer large-scale structures becomes more important due to their larger contribution to skin friction \cite{de2016skin}.
Hence, the efficacy of these near-wall targeted methods decreases with increasing $Re$ \cite{Quadrio2011, gatti2016reynolds, chung2011effectiveness,wang2016active}. 
\citeauthor{SAMIE2020108683} \cite{SAMIE2020108683}  examined the coherence between a  measurable  wall quantity (e.g., the  wall-shear  stress  fluctuations)  and  the  streamwise  and  wall-normal velocity fluctuations in a turbulent boundary layer.
They found that the closed-loop drag reduction scheme targeting near-wall cycle of streaks alone  will be of limited success in practice as $Re$ grows.
Therefore, attempts  have  also  been  made  to  reduce  drag  by  manipulating large-scale  coherent  structures  in  the  logarithmic  and  outer  regions  of  turbulent  boundary  layers.
For example, \citeauthor{schoppa1998large} \cite{schoppa1998large}  developed a conceptually simple
 open-loop  large-scale control strategy.
By imposing large-scale counter-rotating streamwise swirls with a relatively small excitation amplitude,  $20\%$ drag reduction was obtained in a turbulent channel at $Re_\tau \approx 100$.
However, there is an ongoing debate over the effectiveness of the large-scale  control method with increasing $Re$. \citeauthor{canton2016reynolds}\cite{canton2016reynolds} found that the drag reduction decreases with increasing $Re$ and becomes nearly zero at $Re_\tau=550$.
 However, \citeauthor{Jie2017PRF} \citep{Jie2017PRF} showed  that the negative results found by \citeauthor{canton2016reynolds} \cite{canton2016reynolds}
arose from their selection of the channel center height as the fixed value for the location of the control  swirl centers. 
In addition, \citeauthor{Jie2017PRF} \citep{Jie2017PRF}  validated the large-scale drag control concept for high $Re$'s using the  near-wall spanwise opposed wall-jet forcing (SOJF), and a detailed physical explanation on the drag reduction mechanism is furnished in Ref. \red{Note that SOJF introduce weak spanwise friction and its mean is zero.} \cite{yao2018drag}. 
Besides these numerical works, there are  several experimental studies on large-scale drag controls using either jet \cite{iuso2002wall,cannata2020large} or plasma actuation \cite{wong2015active,corke2018active}.

Caution should be given to controlling  large-scale (LSM)
and very large-scale (VLSM) motions, as recent studies show that
suppression of these structures does not yield significant drag reduction.  For example, \citeauthor{abbassi2017skin} \cite{abbassi2017skin} performed an experimental investigation of  closed-loop control of manipulating LSMs and VLSMs, and  only 3.2\% drag reduction was achieved.
This is because, when suppressing these large-scale structures,  smaller scales may also be concomitantly altered due to nonlinear scale interaction \cite{hwang2016inner,de2016skin}.
Hence,  at high $Re$,  the simultaneous  control  of small- and large-scale structures may be required to obtain effective drag reduction.
Therefore,  we propose a composite drag control  (CDC) strategy, where two or more different control methods are combined together and applied simultaneously with  the hope of  producing better performance. Through energy budget analysis,  \citeauthor{chen2021} \cite{chen2021} showed that both OC and SOJF can yield  notable drag reductions, but with different mechanisms.  Their combination  is potentially additive and may even be synergistic -- thus of interest and  is examined in the present work. 
Moreover, there are two additional considerations for performing such a study. First, as OC and  SOJF are effective in suppressing small- and large-scale structures, respectively, their combination may yield a significant drag reduction, even at high $Re$'s.
Second, since both of the two methods require  small power inputs to drive the control,  the net power saving might become promising for CDC.

The rest of the paper is organized as follows.
In the following section, the governing equation, numerical scheme, and control method are described.
The drag reduction results are presented in section \ref{sec:drag}. 
Flow statistics and coherent structures are  examined in section \ref{sec:flow}  to elucidate the underlying mechanism on the variation of skin friction under control. 
Finally, the concluding remarks are provided in section \ref{sec:con}.

% 1. Importance of drag control 

% 2. drag control methods

% 3. issue of drag control 

% 3. composite drag control 

\section{Computational approach}\label{sec:com}

\subsection{Numerical method}
Direct numerical simulations  are performed  in  turbulent channel flows using the code ``POONPACK'' developed by \citeauthor{lee2015direct} \cite{lee2015direct}.
In this study, $x$, $y$, and $z$ denote the streamwise, wall-normal, and spanwise coordinates, respectively; the corresponding velocity components are $u$, $v$, and $w$.
The  incompressible Navier-Stokes   equations are   solved   using   the   method   of  \citeauthor{kim1987turbulence} \cite{kim1987turbulence},   in   which   equations   for   the wall-normal vorticity and the Laplacian of the wall-normal velocity are time-advanced.
This  formulation  has  the  advantage  of  satisfying  the  continuity  constraint  exactly while   eliminating   the   pressure.
A Fourier-Galerkin method is used in the streamwise and spanwise directions, while the  wall-normal  direction  is  represented  using  a  $7$th-order B-spline  collocation  method.
A   low-storage   implicit-explicit   scheme     based  on  third-order  Runge-Kutta  for  the  nonlinear  terms and  Crank-Nicolson  for  the  viscous  terms  are  used  for  time advance.
The flow is driven by a  pressure gradient, which varies in time to ensure that the mass flux  through  the  channel  remains  constant.  For   more   details   about   the   code, refer to  \citeauthor{lee2015direct} \cite{lee2015direct}.

\subsection{Control Schemes}

\begin{figure*}
\centering
\includegraphics[width=0.98\textwidth]{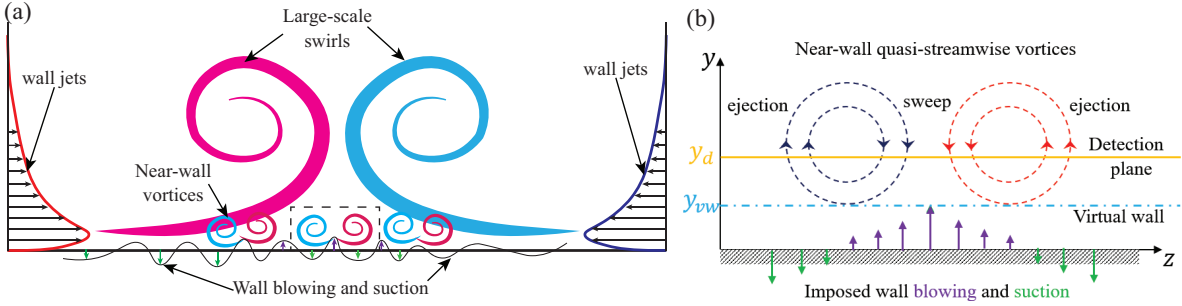}
\caption{(a) Schematic of composite drag control (CDC), which combines the spanwise opposite wall jet forcing (SOJF) and the opposition control (OC);
(b)  the detailed sketch of OC in the near-wall region [dashed box in (a)]. }
\label{fig:Opposition}
\end{figure*}

As mentioned in the previous section, CDC consists of OC and SOJF (Fig.  \ref {fig:Opposition}).
For OC, the wall-normal velocity at the wall is set as
\begin{eqnarray}
 v_w(x,z,t)=-A_o v(x,y_d,z,t),
\end{eqnarray}
where $A_o$ is the control amplitude and $ v(y_d)$ is the  wall-normal velocity at the detection plane $y_d$ (hereinafter, subscript $w$ represents the  value on the wall). Such a wall-velocity setting is imposed on both the bottom and the top walls.

Following \citeauthor{yao2018drag} \cite{yao2018drag}, a spanwise forcing is applied for SOJF:
\begin{eqnarray}\label{eq:ourfor2}
F_z=A_s\sin(\beta z)g(y),
\end{eqnarray}
where $A_s$ is the forcing amplitude, $\beta$ ($=2\pi/\lambda$) is the spanwise wavenumber, and \(g(y^+)\) is a dimensionless forcing as a function of $y$.
Throughout the paper, all control parameters  are  normalized in viscous units based on uncontrolled flow (indicated by the subscript $0$):
a)  $A^+_s=A_s\nu/u^3_{\tau,0}$;
b)  $\lambda^+=2\pi/\beta^+$ ($\beta^+=\beta\nu/u_{\tau,0}$); and
c) $y^+_c=y_cu_{\tau,0}/\nu$.
The forcing function $g(y^+)$ in  Eq. \eqref{eq:ourfor2}
is  $g(y^+)=y^+\exp(-\eta y^{+2})$,
where the decay factor $\eta$ in the wall-normal direction  specifies the wall-jet velocity profile.
The function $g$ has its maximum at $y^{+}=y^+_c=1/\sqrt{2\eta}$, and $y^+_c$ represents
the height of the spanwise wall-jet maximum velocity.
In Eq. \eqref{eq:ourfor2}, $g$ is normalized by its maximum value $g(y^+_c)=y^+_c\exp(-1/2)$.
Note that SOJF is only applied near the bottom wall, but due to its large-scale feature, it is also found to  be effective in reducing drag at the top wall.

\subsection{Simulation parameters}

\begin{table}

    \begin{center}

        \caption{Details of the numerical discretization employed for the present simulations. The computational box size is $4\pi h \times 2h \times 2\pi h$ for all cases,
            and $N_x$, $N_y$ and $N_z$ are the number of grids in x, y, and z, respectively.  }{\label{tbl:Num}}
        \begin{tabular}{ccccccc}
            \hline
            $Re_\tau$ & $Re_b$  & $N_x\times N_y\times N_z$&$\Delta x^+$&$\Delta y^+$&$\Delta z^+$\\
            \hline
            $180$& $2857$ & $256 \times 192\times 256$& $9.8$&$0.11-4.7$&$4.9$\\
            \hline
        \end{tabular}
    \end{center}
\end{table}

DNS  is conducted at a fixed bulk Reynolds number $Re_b=U_bh/\nu$, where $U_b$ is the bulk velocity, and the corresponding $Re_\tau \approx180$.
The domain sizes, grid sizes, and resolutions used in the  simulations are shown in Table \ref{tbl:Num}.
The adequacy of the domain size and resolution is confirmed in Refs. \cite{yao2018drag,chen2021}.
There are, respectively, three and two control parameters for SOJF and OC --  a  total of five parameters, which make the search for optimal control rather challenging. 
For simplicity, here we  consider only one critical control parameter for each method, namely, $A_s$ for SOJF, and $y_d$ for OC. For other control parameters, we choose the optimal values reported in previous works: namely, $\lambda=2\pi$ (or, equivalently, $\lambda^+\approx 1000$) for simultaneous control of about ten streaks and $y^+_c=30$ corresponding to the peak of Reynolds shear stress \citep{yao2018drag}, and $A_o=1$ \cite{choi1994active,hammond1998observed}. 
In the current study, $A^+_s$ varies between $[0,2/3,4/3,2,8/3]\times 10^{-3}$ and $y^+_d=[0,5,10,15,20]$.
All computations start with the same initial condition of a fully developed turbulent channel flow without control.
Time  averaging for statistics is taken  over
at  least  25 eddy turnover time (i.e., $tu_\tau/h\ge25$)  after  the  initial  transition.

% All simulations are carried out at a constant time-marching step $\Delta t^+ = 0.1$, with  
% 

Due to the presence of large-scale coherent swirls induced by the SOJF, the total flow field can be divided into 
three components by using the triple decomposition \citep{reynolds1972mechanics}:
\begin{eqnarray}\label{eq:trp}
 \mathbf{u}(x,y,z,t)= \mathbf{U}(y)+\underbrace{\tilde{\mathbf{u}}(y,z)+\mathbf{u}''(x,y,z,t)}_{\mathbf{u}'}.
\end{eqnarray}
Here, $\mathbf{u}$ is the instantaneous total velocity field; $\mathbf{U}=\langle \overline{\mathbf{u}}\rangle$
is the mean velocity, where the overbar indicates averaging in time and in the streamwise direction;
and the bracket $\langle\cdot\rangle$ indicates spanwise  averaging over one wavelength period;
$\mathbf{u}'=\mathbf{u}-\mathbf{U}$ is the total fluctuation;
$\tilde{\mathbf{u}}=\overline {\mathbf{u}}-\mathbf{U}$ is the so-called ``organized field''
representing the coherent motion induced by the swirls of SOJF, and $\mathbf{u}''=\mathbf{u}-\overline {\mathbf{u}}$ is the random turbulent fluctuation.
Note that for OC only, $\tilde{\mathbf{u}}$ is absent, and $\mathbf{u}''=\mathbf{u}'$.

\section{Drag reduction and control efficiency} \label{sec:drag}

% \subsection{Drag reduction }

We define drag reduction as the relative change in  the skin-friction drag coefficient:
\begin{eqnarray}
\mathcal{R}=1-C_f/C_{f,0},
\end{eqnarray}
where $C_{f}=2\tau_w/(\rho_b U^2_b)$ and  $C_{f,0}=2\tau_{w,0}/(\rho_b U^2_b)$ are the drag coefficients of the controlled and reference (uncontrolled) cases, respectively.

% \textcolor{blue}{In the current study, $C_f$ is calculated based on the wall-shear stress of both top and bottom walls ????}

Net power saving can be defined as a relative change  in  the power spent:
\begin{eqnarray}
\mathcal{N}=1-(P+W)/P_{0},
\end{eqnarray}
where $P=u^2_\tau U_b/h$ and  $P_0=u^2_{\tau,0} U_b/h$ are the pumping power of the controlled and reference (uncontrolled) cases, respectively; and $W$ is the power input due to control, which is given as \cite{chen2021}
\red{
\begin{equation}\label{eq:Wcontrol}
W=\underbrace{\langle{\overline{wF_z}}\rangle}_{W_s}+\underbrace{\langle \overline{\rho v^3_w/2}\rangle+\langle{\overline{p_w v_w}}\rangle}_{W_o}.
\end{equation}
Here,  $F_z$ is the spanwise force  given in Eq. \ref{eq:ourfor2}, and $p_w$ and $v_w$ represent pressure and wall-normal velocity at the wall. $W_s$ and $W_o$ denote the power inputs to the fluid system by  SOJF and OC, respectively.}
For  statistical equilibrium or steady state, the total power input is balanced by total dissipation (i.e., $\epsilon=\nu\langle\overline{\partial_j u_i\partial_j u_i}\rangle$).  Hence, the  net power saving rate  can also be estimated as
\begin{equation}\label{eqn:dragNeta}
\mathcal{N}=1-\epsilon/\epsilon_0,
\end{equation}
where $\epsilon$ and  $\epsilon_0$ denote the dissipation of the controlled and reference (uncontrolled) cases, respectively.

Note that Eq. \ref{eq:Wcontrol} is  the exact mathematical form of the control power input \cite{chen2021}. 
Defined this way, $W$ can have locally  negative values. 
To ensure  that any power input is always associated with energy consumption,  several works \cite{stroh2015comparison,deng2014strengthened} employ a more conservative form to estimate the power input by taking the absolute values of these quantities  before averaging.
However, in this work, we use this exact form, as it reflects the true energy flux in the flow systems (see Sec.\ref{sec:energ}).

% and also provides a convenient way to estimate the net power saving from the subsequent Eq. \ref{eqn:dragNeta}

\begin{figure*}[ht!]
\centering
\centering
       \subfloat[]{
          \label{fig:DR1:a}
      \includegraphics[width=0.48\textwidth]{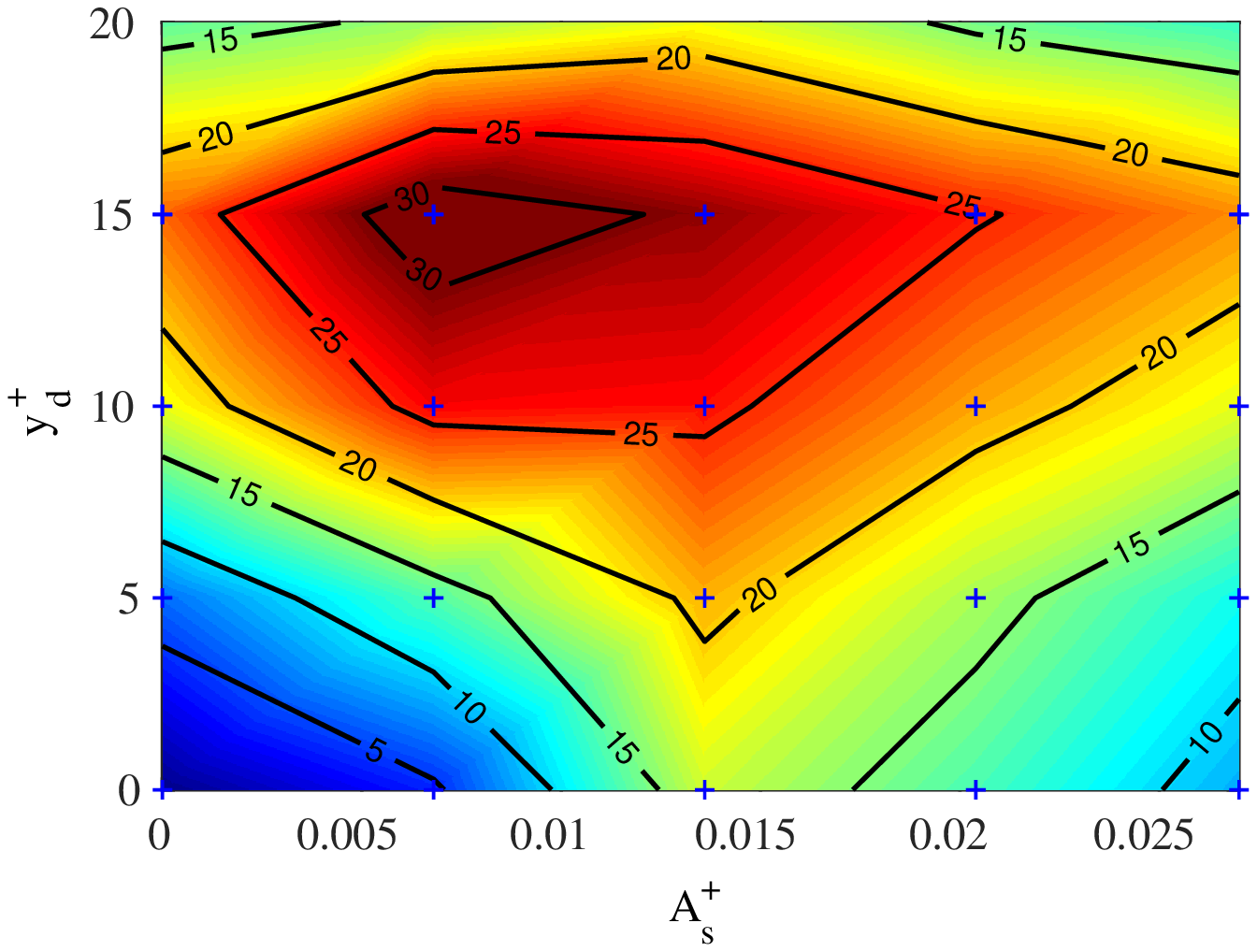}}
      \subfloat[]{
          \label{fig:DR1:b}
      \includegraphics[width=0.48\textwidth]{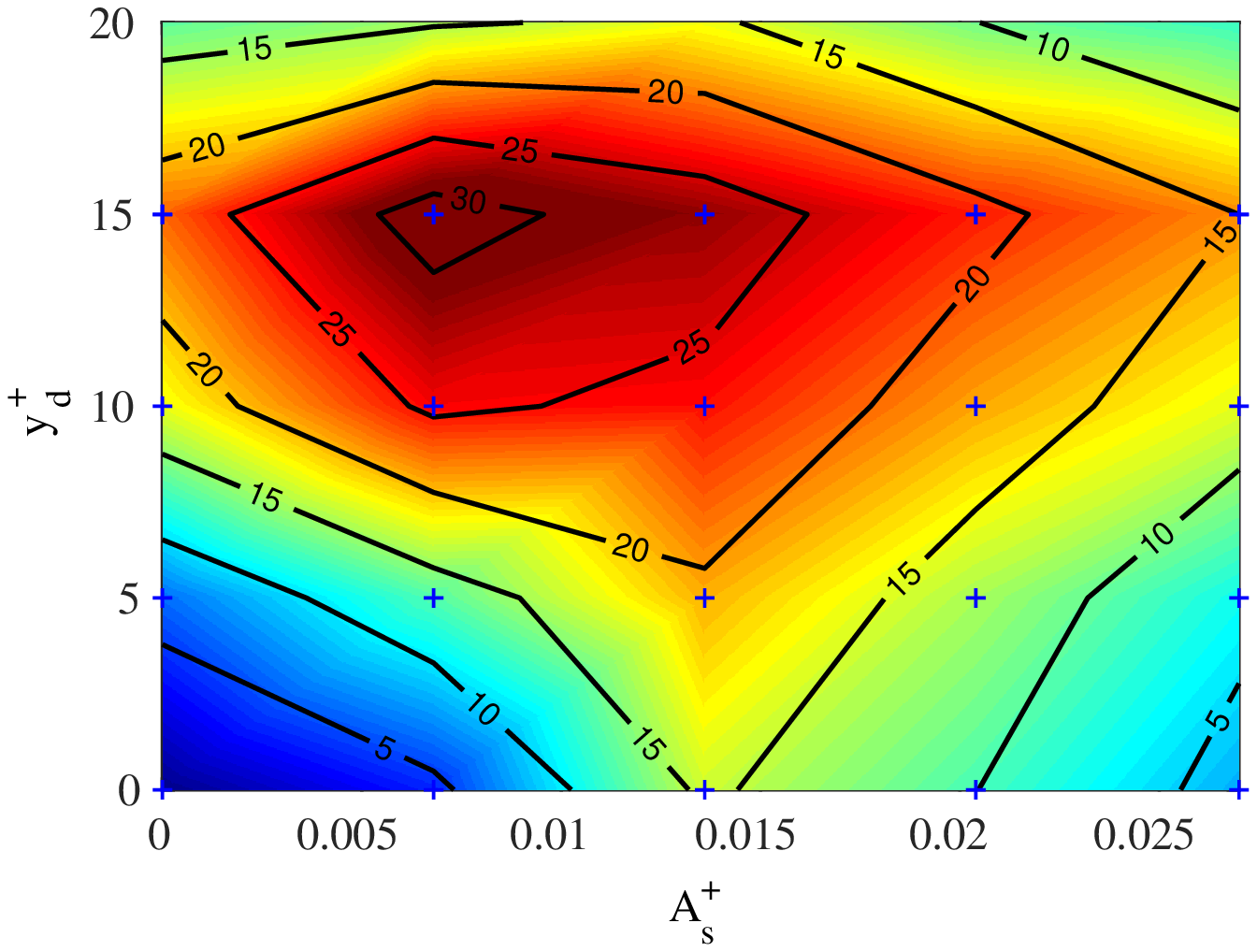}}
\caption{(a) Drag reduction  $\mathcal{R}$ ($\%$) and (b) net power saving $\mathcal{N}(\%)$ as  functions of $A^+_s$ and $y^+_d$ for CDC at $Re_\tau=180$.
The symbol $+$ indicates simulation data points used for interpolation.}
\label{fig:DR}
\end{figure*}

\begin{figure*}[ht]
\centering
    \subfloat[]{
          \label{fig:DR2:b}
      \includegraphics[width=0.45\textwidth]{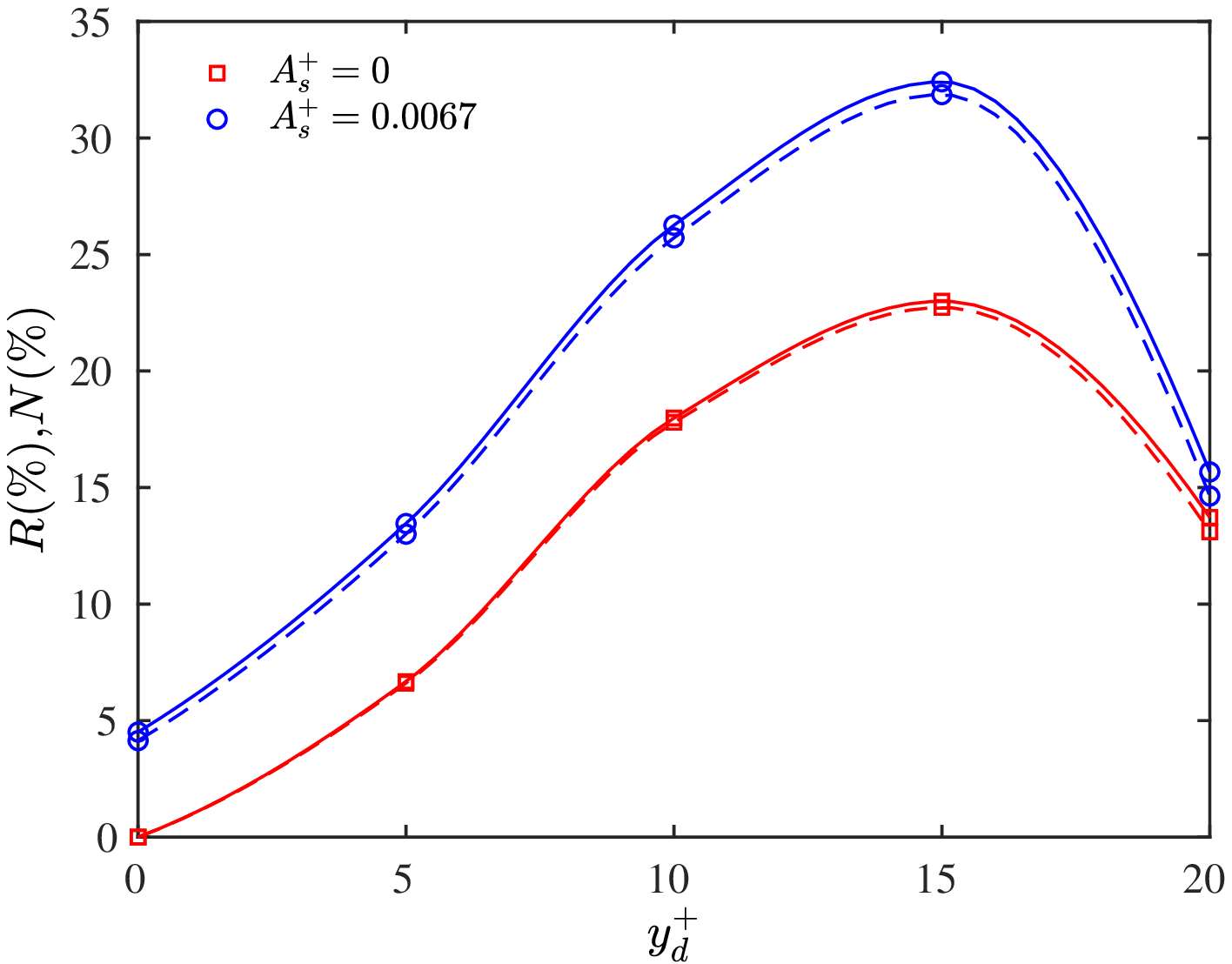}}
       \subfloat[]{
          \label{fig:DR2:a}
      \includegraphics[width=0.45\textwidth]{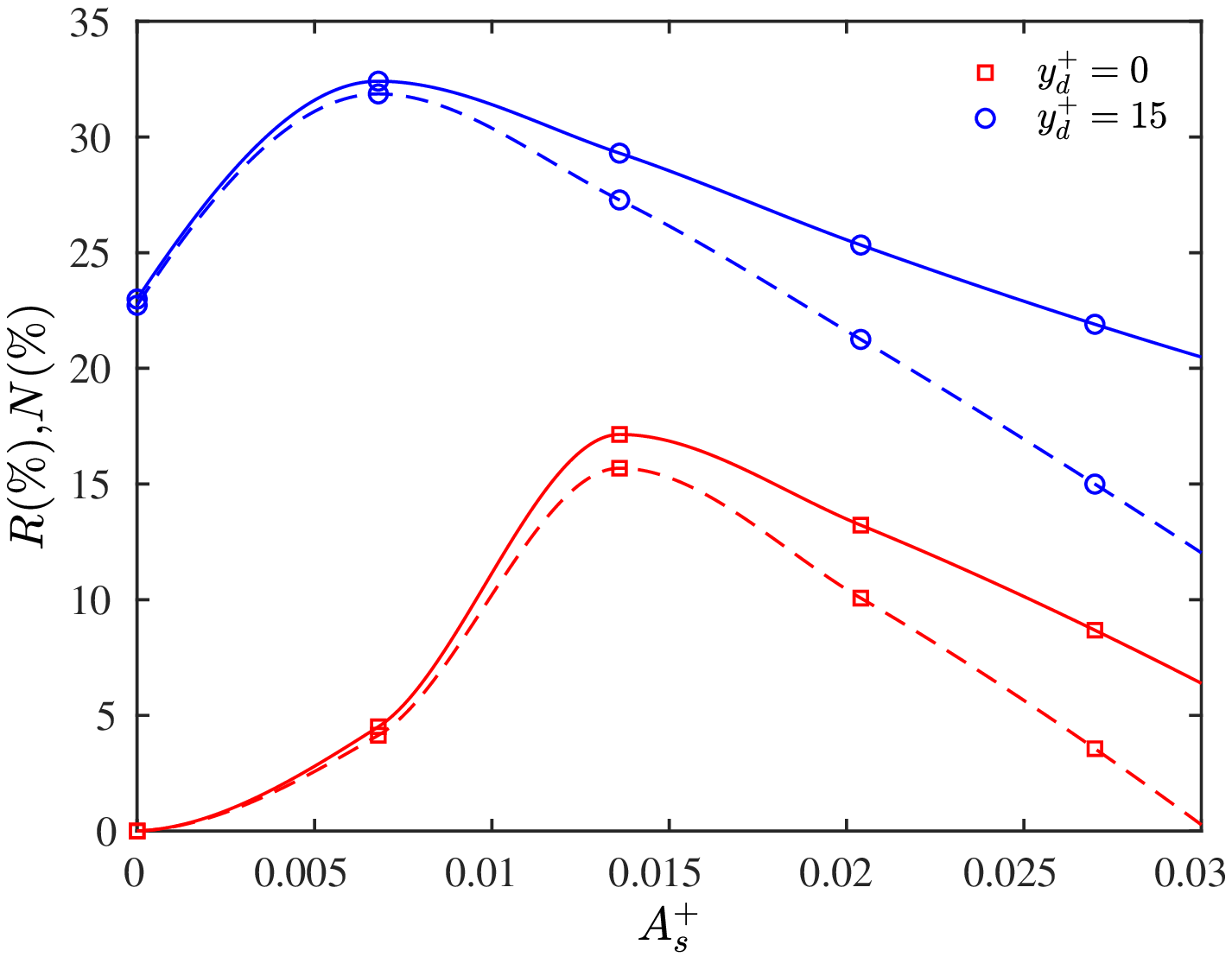}}
      
\caption{Drag reduction $\mathcal{R}(\%)$ (solid lines)  and net power saving $\mathcal{N}(\%)$ (dashed lines) as a function of (a)  $y^+_d$ and (b) $A^+_s$.}
\label{fig:DR2}
\end{figure*}

Figures \ref{fig:DR} (a,b) show the contour plots of drag reduction $\mathcal{R}$ and net power saving $\mathcal{N}$ as  functions of $A^+_s$ and $y^+_d$ at $Re_\tau=180$.
The contours are computed from a linear interpolation of   discrete simulated points (marked as \textcolor{blue}{$+$}).
Note that  cases with $y^+_d=0$ ($A^+_s=0$) corresponds to the SOJF (OC) only. 
In addition, figures \ref{fig:DR2}(a,b)  show the variation of $\mathcal{R}$ and $\mathcal{N}$ as  functions of  $A^+_s$ and $y^+_d$, respectively.
It is clear from Fig. \ref{fig:DR} (a) that significant $\mathcal{R}$  can be achieved in broad ranges of  $A^+_s$ and $y^+_d$. 
For a given $A^+_s$, $\mathcal{R}$ first increases then decreases with $y^+_d$, and  the maximum $\mathcal{R}$ is always obtained at $y^+_d=15$ -- the optimal value for OC only. 
For a given $y^+_d$, $\mathcal{R}$ also first increases then decreases with $A^+_s$.
However,  different from $y^+_d$, the optimal $A^+_s$ changes and  depends on $y^+_d$. 
When $y^+_d=0$, the maximum $\mathcal{R}$ is achieved at $A^+_s=0.0013$, which is very close to that reported in Ref. \cite{yao2018drag}. As $y^+_d$ increases, the value of $A^+_s$ that produces maximum $\mathcal{R}$ first decreases and then increases. 
\red{Within the parameters considered, the maximum $\mathcal{R}$ is $32.5\%$, with $y^+_d=15$ and $A^+_s=0.0067$, 
 and the uncertainty,  estimated by following the procedure in \cite{gatti2016reynolds,yao2018drag}, is approximately $0.5\%$.}
Note that as only the variations of $y^+_d$ and $A^+_s$ are considered  here, a higher $\mathcal{R}$ could possibly be found by varying other control parameters. This is not pursued here due to  limited computational resources.
\red{It is also worth mentioning that the current studies are performed under the constant flow rate assumption, and,  
at this low $Re$, the actually drag reduction amount would be slightly different if the constant pressure gradient assumption is employed \cite{gatti2016reynolds}. However, the overall  behavior of  $\mathcal{R}$ as shown in Fig. \ref{fig:DR}a should remain the same. }
%  however, the overall drag reduction behavior with respect to the control parameters would not be affected.
% ; however, the composite control idea is clearly shown to be effective in our present study.

% tion, which is typically of the order of 100 ∼
% 300h/U b as shown in Yao et al. (2017). For the drag reduction rate R , 

% \subsection{Net power saving}

% \begin{equation}\label{eq:Wcontrol}
% W= {\langle{u_iF_i}\rangle}+{{J_w}},\quad J_w=\langle v_wv_{w}v_{w}\rangle/2H+\langle{p_w v_w}\rangle/H.
% \end{equation}
% Here, ${\langle{u_iF_i}\rangle}$ is energy spent by external force, and $J_w$ is the energy input due to OC control. In the above expressions, $\langle{\phi}\rangle_{t,x,y,z}=\int^T_0\int^{L_x}_0\int^{L_z}_0\int^{2H}_0\frac{\phi}{{2HL_xL_zT}} dydxdzdt$ indicats time and volume integration; subscript $w$ indicates value at the wall.

\citeauthor{chen2021} \cite{chen2021} showed that compared to other active control techniques, such as spanwise wall oscillation,   both OC and SOJF are very promising for net power saving. 
For OC, the power input $W_o$ increases as $y^+_d$ increases due to  high intensity of wall-normal velocity fluctuations \cite{deng2014strengthened}; for SOJF,  $W_s$ increases with $A^+_s$ due to stronger large-scale swirls \cite{yao2018drag}.
However, as both SOJF and OC controls require little power input, notable net power saving can be achieved when combining them together. As shown in Fig. \ref{fig:DR},
the contour plot of $\mathcal{N}$ is rather similar to those of $\mathcal{R}$, particularly for small $y^+_d$ and $A^+_s$.
Fig. \ref{fig:DR2}  further shows that for a given $A^+_s$,  $\mathcal{N}$ is only slightly smaller than  $\mathcal{R}$, and for a given $y^+_d$, $\mathcal{N}$ slowly deviates from $\mathcal{R}$ with increasing  $A^+_s$. For the optimally controlled case (i.e., $y^+_d=15$ and $A^+_s=0.0067$),  approximately 1\% of pumping power  is needed to actuate the flow control, which  results in a maximum $\mathcal{N}$ of $31.5\%$.
\redsec{Although this high net power saving is impressive, some caution is in order  as this value is  based only on a
simple estimation of the actuation, where the actuator’s energy transfer efficiency (e.g., that of plasma actuators) has not been taken into consideration.}

In summary, with the limited parameter search, we demonstrate that the CDC technique could yield a maximum of $\mathcal{R}=32.5\%$ and $\mathcal{N}=31.5\%$ -- much higher than that can be achieved by either OC or SOJF. This is indeed a very   promising finding.
% Thus, a significant $\mathcal{N}$ is also achieved with a broad range of $A^+_s$ and $y^+_d$. 

\section{Flow analysis} \label{sec:flow}

To understand the mechanisms of drag reduction,
flow statistics as well as the dynamics are examined in detail below for four cases: Case I, $A^+_s=0$ and $y^+_d=0$ (uncontrolled); Case II, $A^+_s=0$ and $y^+_d=15$ (optimal for OC only);  Case III, $A^+_s=0.0133$ and $y^+_d=0$ (optimal for SOJF only); and Case IV, $A^+_s=0.0067$ and $y^+_d=15$ (optimal for CDC).

 \subsection{\red{Near-wall streaks and vortical structures}}

\begin{figure*}
\centering
\includegraphics[width=0.95\textwidth]{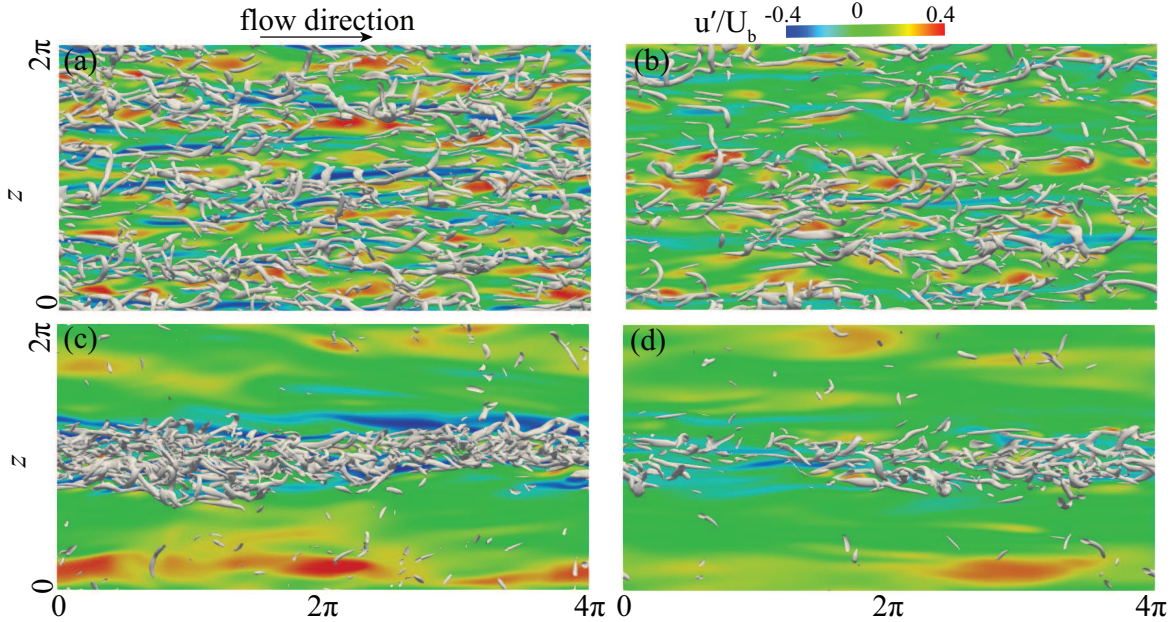}
\caption{ $\lambda_2$ vortical structures  associated with the streamwise velocity fluctuations at $y/h=-0.92 (y^+_0=15)$ on the bottom-half walls for (a) Case I; (b) Case II; (c) Case III; and (d) Case IV.}
\label{fig:cs}
\end{figure*}

Figure \ref{fig:cs} shows the $\lambda_2$ vortices \cite{jeong1995identification} along with the low-speed streaks at $y^+=15$.
For Case I, the  typical meandering low-speed streaky structures are observed aligned along the streamwise direction, and 
numerous slender vortices are distributed throughout  the wall region,
lying around the interface between low- and high-speed streaks [figure \ref{fig:cs}(a)].
For Case II, although the streaks still have  typical meandering shapes, their strengths are significantly weakened.
Correspondingly, the generation of drag-inducing near-wall
streamwise vortices is also suppressed, with fewer vortices as expected. 
Under SOJF, the basic near-wall streaks are destroyed [Figs. \ref{fig:cs}(c,d)], and only a single wide meandering streak lies along the mid-span and extends the entire flow length, as expected from the merger of the numerous background streaks forced by the two wall-jets \cite{yao2018drag}.
For Case III, most of the near-wall streamwise vortices disappear, except near the central collision region, just above  the merged streak envelope.
Consistent with the further suppression of  random turbulent fluctuations and  associated dissipation, the  vortices become fewer for Case IV -- clarifying the improved effectiveness of CDC. 
 With OC applied, the near-wall streaks become weak; hence less intensive large-scale swirls are required to merge them together -- the 
reason why the optimal $A^+_s$  decreases from $0.015$ to $0.0067$ [Fig. \ref{fig:DR2}(b)]. 
% Interestingly, , while the optimal detecting plane $y^+_d$ of OC  does not change. 
% The reason is perhaps that: for SOJF control, $A^+_s$ represents the velocity intensity of large-scale swirling motion that collects and merges low-speed streaks together. 
% In contrast, although SOJF control leads to a lower-flank angle of streak envelope and less streamwise vortex generation, it does not change the bursting and sweeping motions associated with the streamwise vortices. Therefore, the optimal detecting plane for OC  is the same as  $y^+_d=15$ no matter SOJF is added or not. 

\subsection{Flow Statistics}

\begin{figure*}
\centering
       \subfloat[]{
      \includegraphics[width=0.33\textwidth]{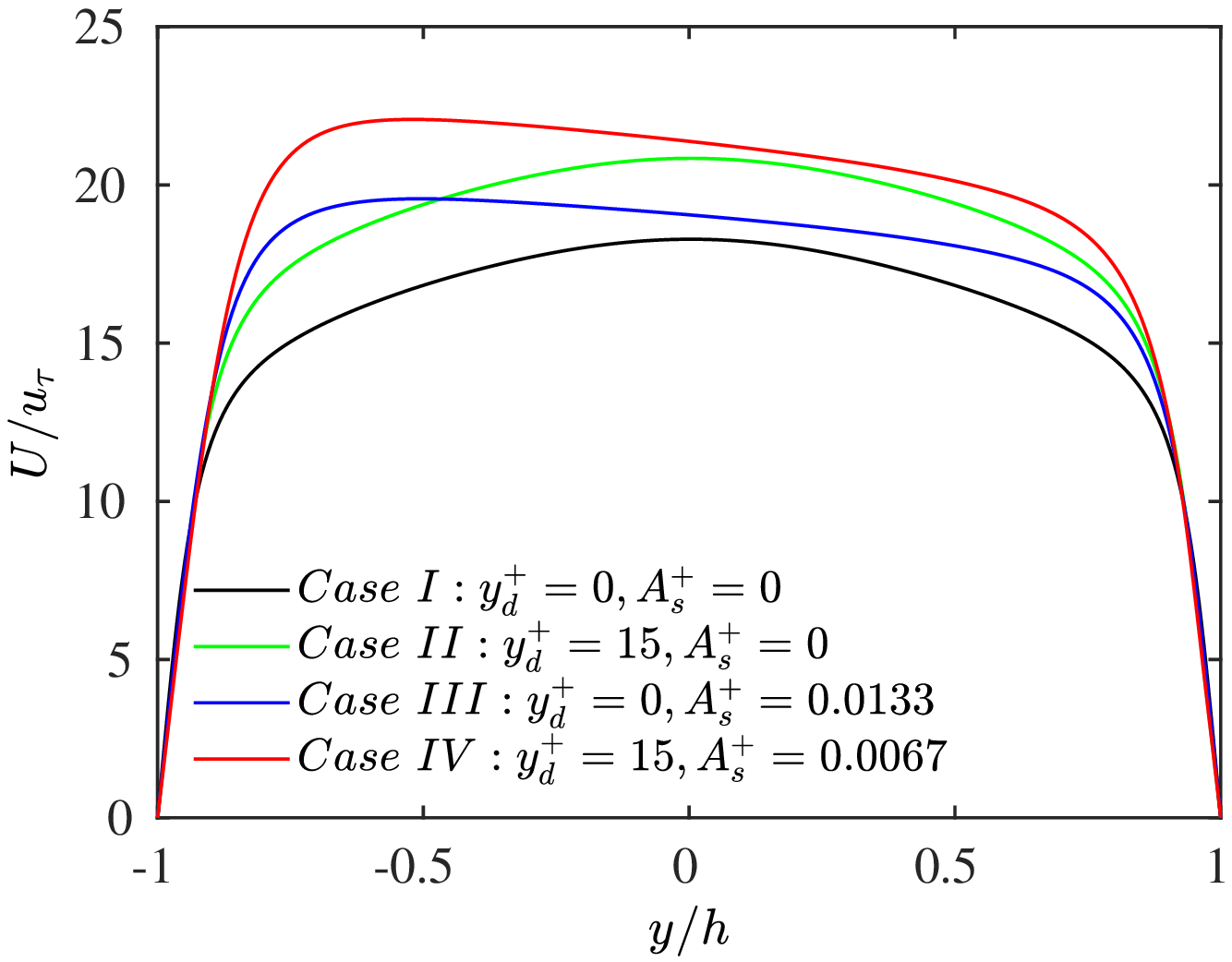}}
      \subfloat[]{
      \includegraphics[width=0.33\textwidth]{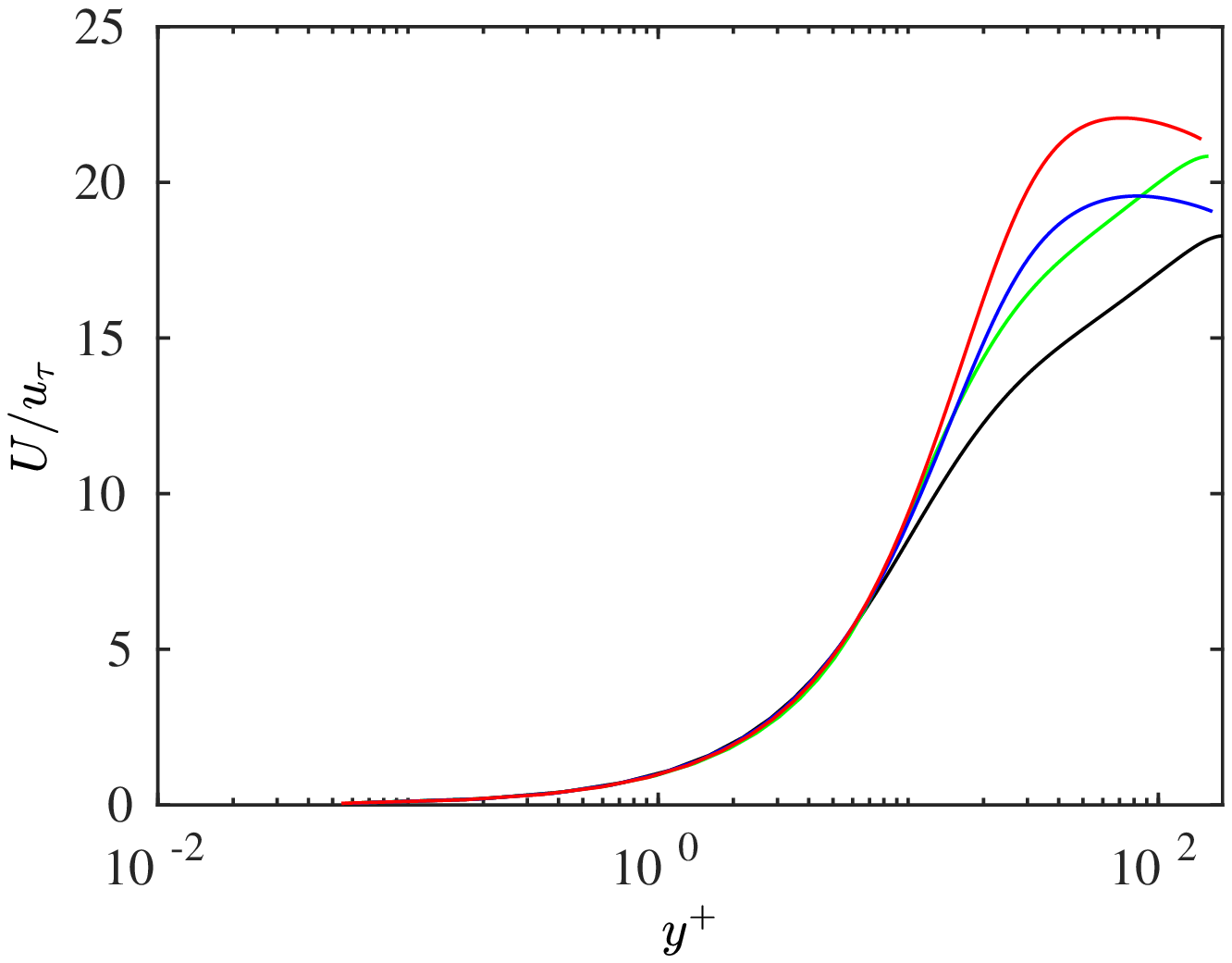}}
      \subfloat[]{
      \includegraphics[width=0.33\textwidth]{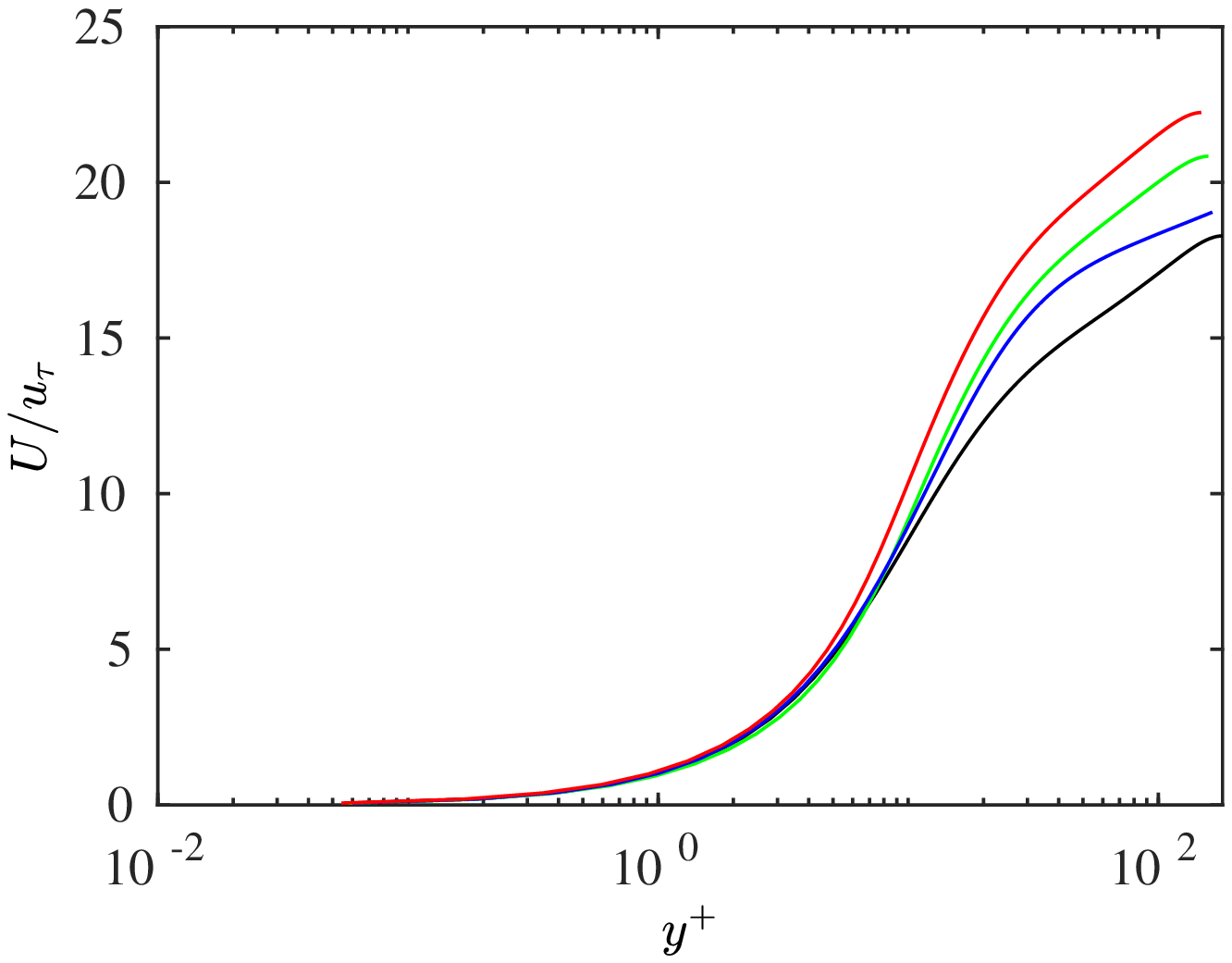}}
\caption{Mean velocity $U(y)$ for (a) the whole channel, (b) bottom- and (c) top-half channel. }
\label{fig:mean}
\end{figure*}

Figure \ref{fig:mean}(a) shows the mean streamwise velocity profile $U^+ (=U/u_\tau)$ in outer units, with Figs. \ref{fig:mean}(b) and \ref{fig:mean}(c) showing the same quantity for the bottom and top halves of the channel in the  wall units. 
Compared to the uncontrolled case (i.e., Case I),  $U^+$ profile for Case II (corresponding to the optimal OC case) in the near-wall region is almost unchanged, and $U^+$ in the log and  outer region shows a characteristic upward shift  --  similar to those observed in Refs. \cite{chung2011effectiveness,deng2014strengthened}. 
As the flow in the outer region is barely altered by the OC, the upward shift is mainly attributed to a smaller $u_\tau$; hence,  its magnitude is directly linked to   $\mathcal{R}$.
For Case III (corresponding to  the optimal SOJF control case),   $U^+$ profile is significantly altered. 
In particular,   due to the presence of large swirls imposed by SOJF near the bottom wall, $U^+$ is  no longer symmetric with respect to the channel centerline, and the peak of  $U^+$  is shifted towards the bottom wall. For Case IV (corresponding to the optimal CDC case), $U^+$ resembles  Case III, but with an upward shift. 
% This is clearly due to the increased wall-normal induced by the vortices in the channel centre region

\begin{figure*}[ht]
\centering
       \subfloat[]{
      \includegraphics[width=0.48\textwidth]{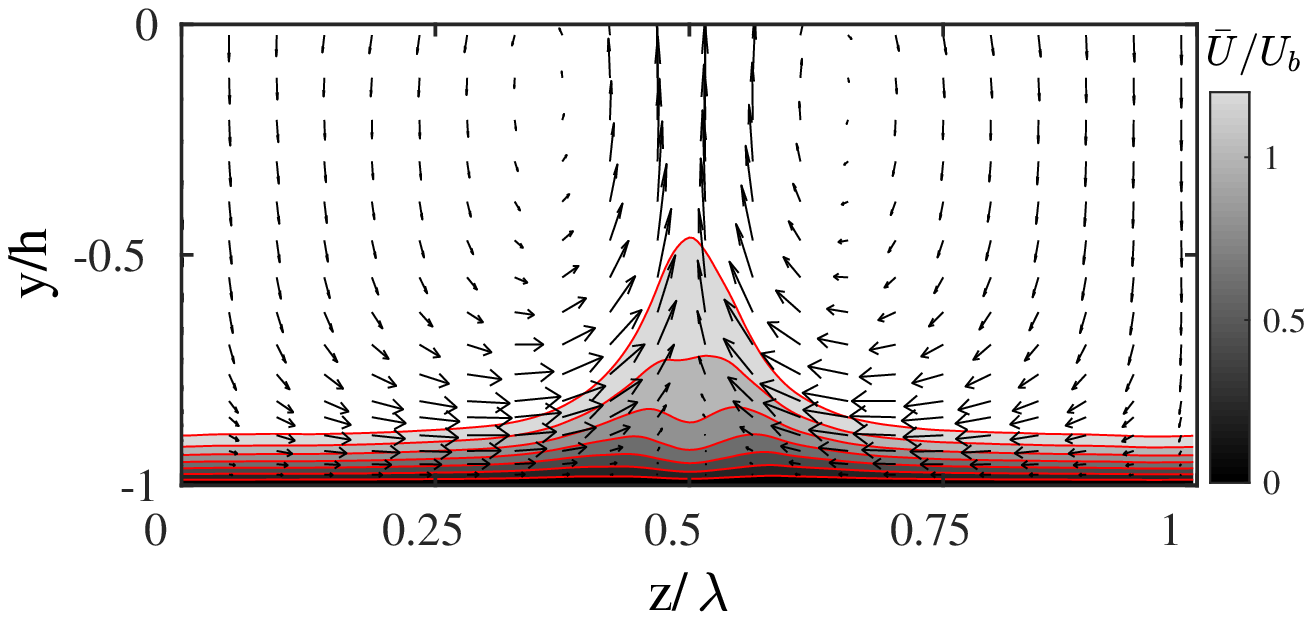}}
      \subfloat[]{
      \includegraphics[width=0.48\textwidth]{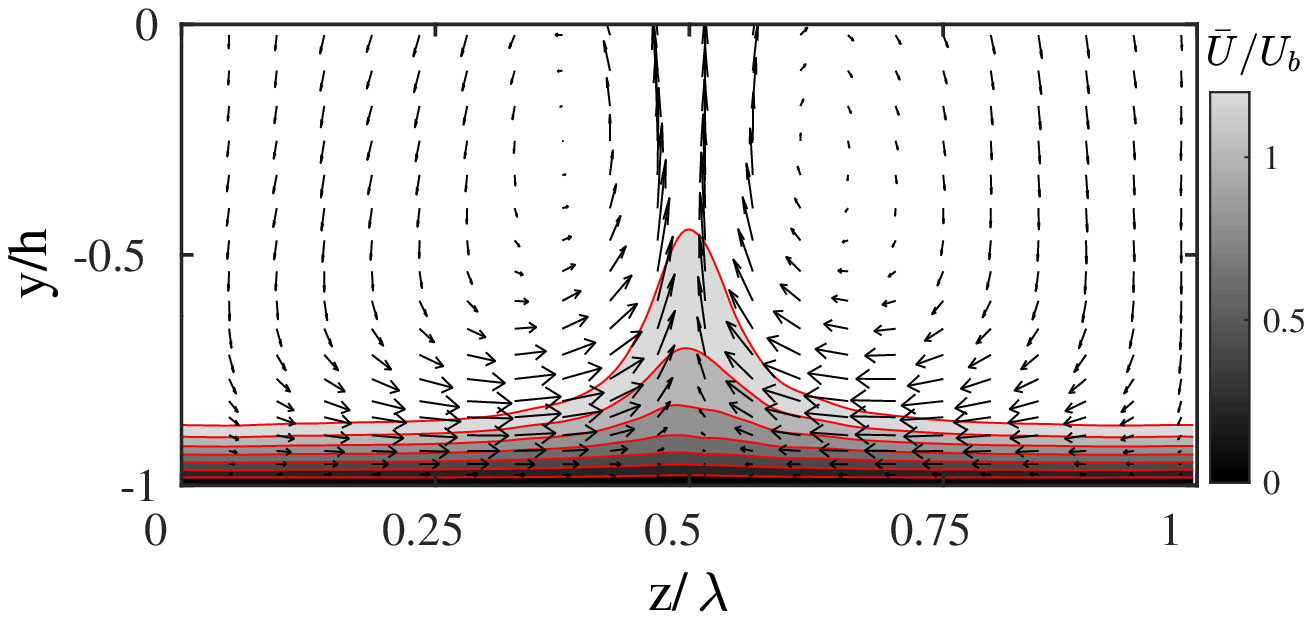}}
\caption{Cross-stream ($y-z$) plane  view of the mean velocity field ($\overline{\mathbf{u}}$)
        with  the contours indicating the mean streamwise velocity (normalized by the bulk velocity $U_b$), and
        the vectors denote the wall-normal and spanwise velocities for (a) Case III and  (b) Case IV.  }
\label{fig:mean2d}
\end{figure*}

Figures \ref{fig:mean2d}(a) and \ref{fig:mean2d}(b) show the cross-streamwise view ($y-z$ plane) of
contours of the streamwise velocity and the cross-plane velocity vectors of
$\overline{ \mathbf{u}}$ for Case III 
and  Case IV, respectively.
As discussed in Ref. \cite{yao2018drag}, due to  the lift-up effect
of the counter-rotating control swirls, a low-speed streak  forms for Case III and extends up to the outer flow region ($y/h>-0.5$).
Near the wall, however, the streaks have dual peaks -- due to the secondary swirls caused by the no-slip wall.
As for Case IV, the topmost low-speed streak has a similar shape as Case III. However, due to the presence of  OC, the secondary swirls are suppressed; as a result,  the dual peaks of the  near-wall streaks become less apparent.
In addition, due to smaller $A^+_s$, the primary counter-rotating swirls are also weaker than that of Case III.

\begin{figure*}
\centering
       \subfloat[]{
      \includegraphics[width=0.33\textwidth]{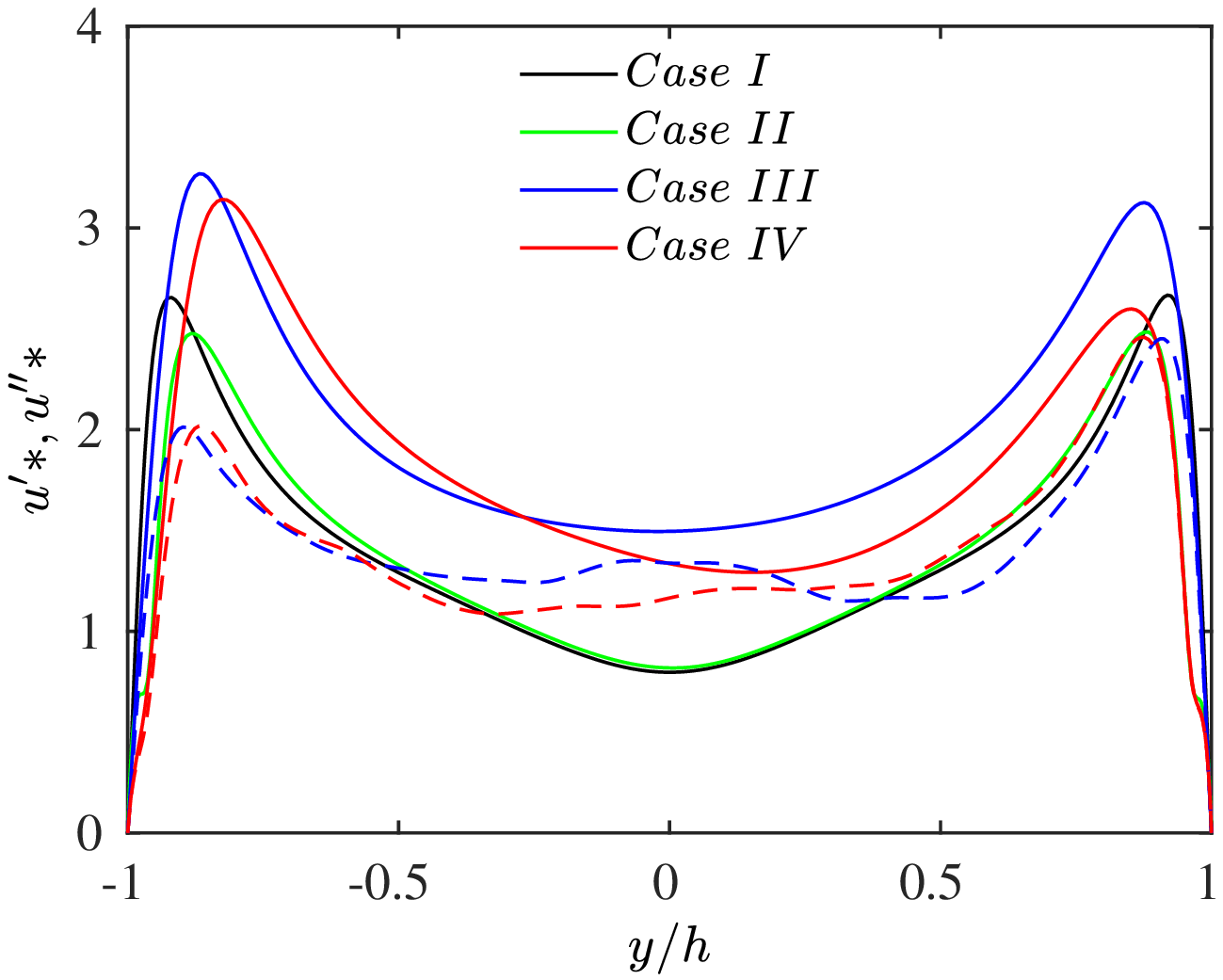}}
      \subfloat[]{
      \includegraphics[width=0.33\textwidth]{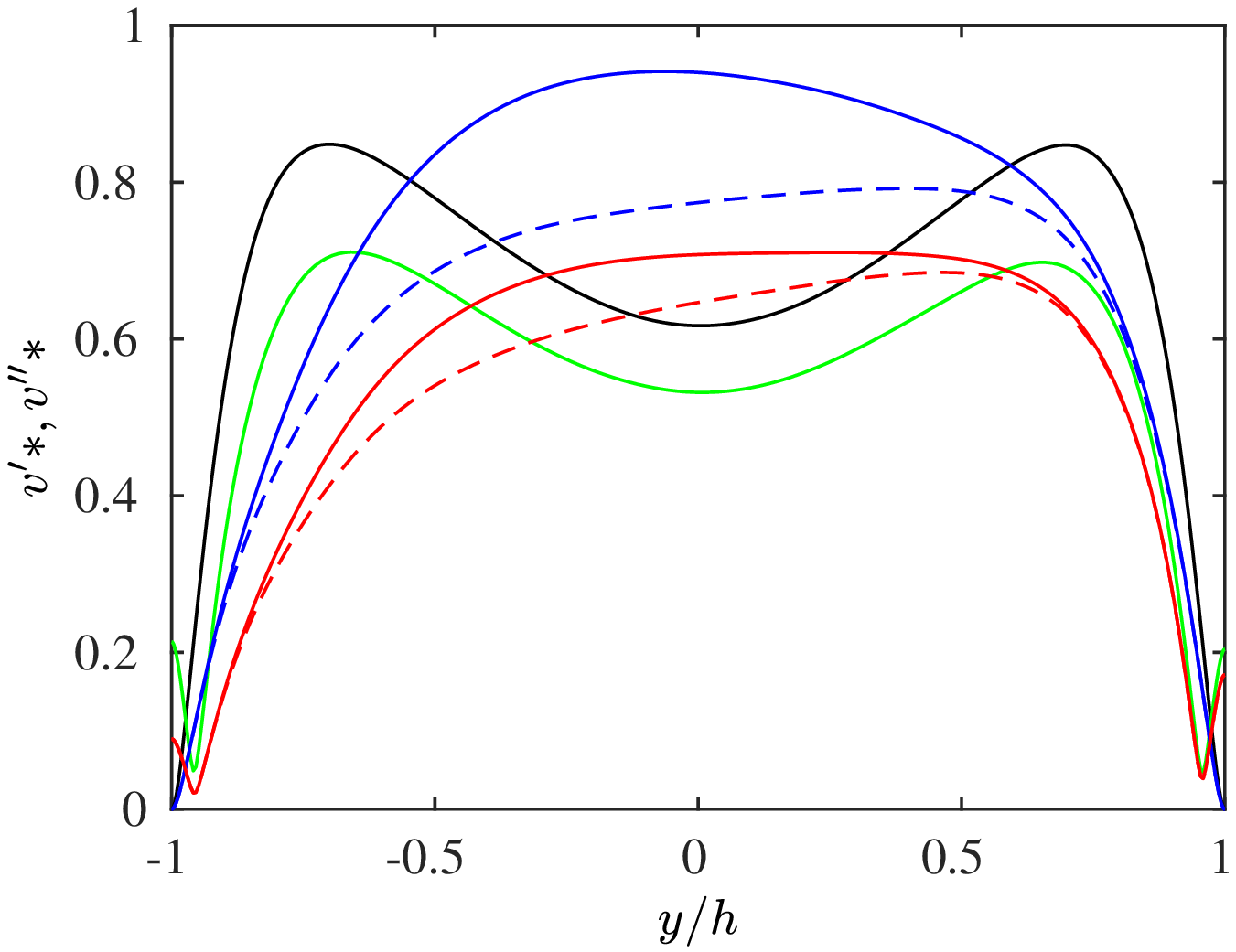}}
      \subfloat[]{
      \includegraphics[width=0.33\textwidth]{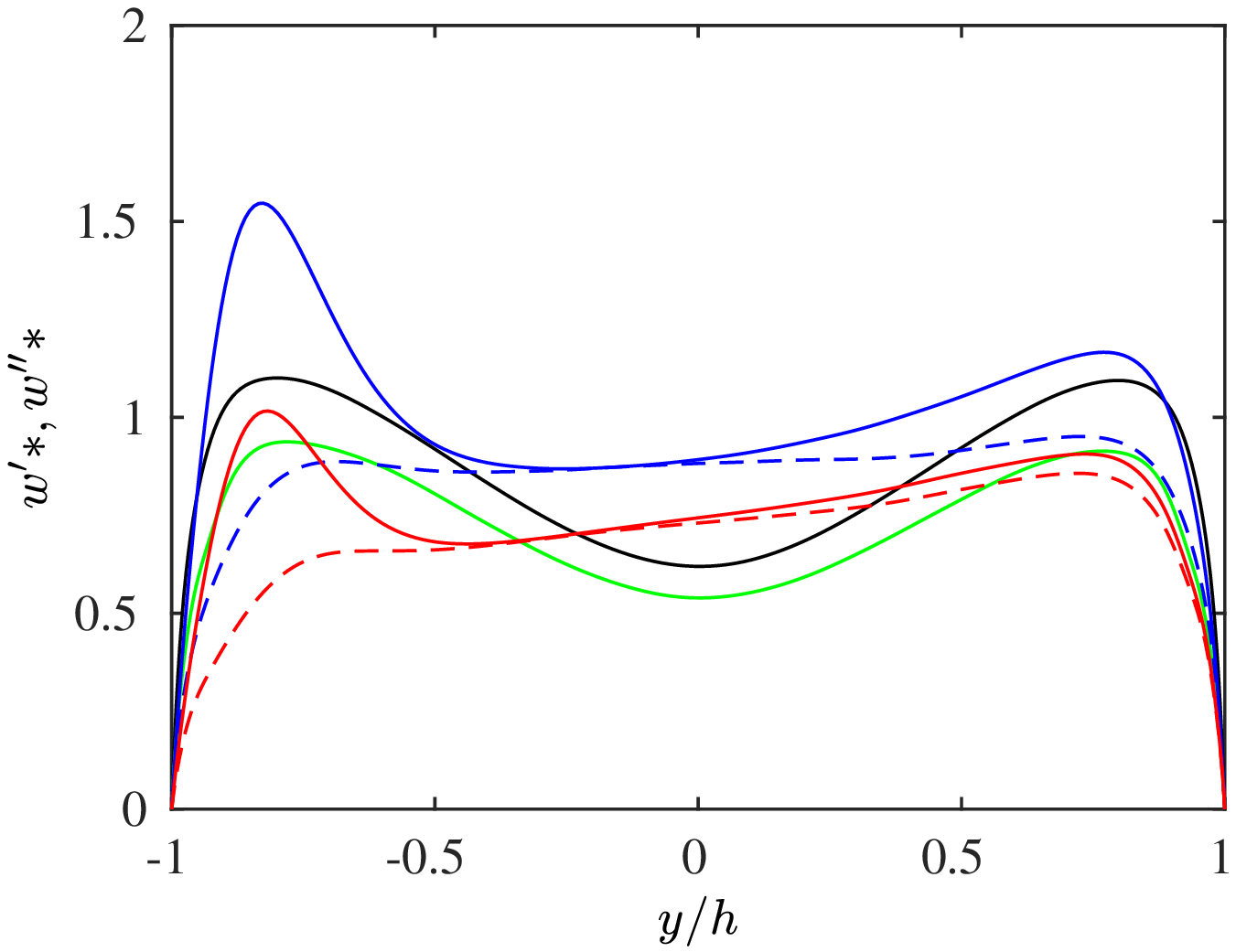}}
\caption{Root mean square of velocity fluctuations: (a) $u'_{rms}$;  (b) $v'_{rms}$;  (c) $w'_{rms}$ . Note that the dashed lines are $u''_{rms}$ and $v''_{rms}$ and $w''_{rms}$ for Cases III and IV.    }
\label{fig:rms}
\end{figure*}

Figure \ref{fig:rms} shows the root mean square (r.m.s) of  total velocity fluctuations  ($u'^*$, $v'^*$, and $w'^*$) and random velocity fluctuations  ($u''^*$, $v''^*$, and $w''^*$). [The superscript $*$ indicates non-dimensionalization by the wall units of the controlled flow]. 
For Case II,  OC establishes a virtual wall as illustrated in Fig. \ref{fig:Opposition}b, which  is typically located halfway between the physical wall and the sensing plane $y^+_d$ \cite{choi1994active,chung2011effectiveness}. 
The location of the virtual wall $y_{vw}$ can  be reflected by a local minimum of $v'^*_{rms}$.
Turbulent intensities are significantly suppressed
near the virtual wall.
In particular, $v'_{rms}$  becomes almost zero  at $y=y_{vw}$, which inhabits the interaction between the near wall and the outer core regions. 
As a result, $u'_{rms}$ and $w'_{rms}$ are significantly suppressed below the virtual wall. 
Compared to the uncontrolled case, the peak location moves away from the wall, consistent with the observation in Refs. \citep{choi1994active,deng2014strengthened}.
For Case III, although all turbulence intensities ($u'_{rms}$, $v'_{rms}$, and $w'_{rms}$) are suppressed in the near-wall region, they are significantly enhanced in the outer region -- mainly due to the increased  momentum exchange caused by the presence of large-scale control swirls. 
For random turbulence velocity fluctuations ($u''_{rms}$, $v''_{rms}$, and $w''_{rms}$), where the effects of large-scale swirls are excluded,  the peaks of them become smaller than those of the uncontrolled case. 
However, they are still slightly larger than the uncontrolled case in the core layer, mainly due to the advection of the large-scale control swirls, which transport the near-wall turbulence and vortical structures to the outer flow.
For Case IV, the behaviors of all turbulence intensities are similar to  Case III, but with smaller magnitudes -- suggesting a better suppression of near-wall streamwise vortices. 
%  due to the presence of , the peaks of $u'_{rms}$, $v'_{rms}$, and $w'_{rms}$  are increased. 

% although the turbulent intensity are suppressed very near the wall.
% In addition, they are all enhanced in the outer region, 
% For all the cases, the velocity  fluctuations are suppressed in the near-wall region.

% owever,
%  (see § 4.3).

% % the virtual wall $y_{vw}$ roughly occurs when $\rho v$ is minimal
% 
% 
% Due to the actuation applied at the wall, namely, opposite blowing/suction,
% the wall-normal Reynolds stress $\tau_{22}$ and shear stress $\tau_{12}$ are enhanced very close to the wall. 
% For the optimally controlled case, $\tau_{12}/\tau_{w,0}$ on the wall is $0.049$ and $0.0428$ for $M_b=0.8$ and $M_b=1.5$, respectively, which are slightly smaller than the value of $0.0625$  reported in Ref. \cite{chung2011effectiveness} for the strictly incompressible case. 
% When the sensing plane is further away from the wall,  $\tau_{12}$ below the virtual wall  significantly increases,  which is  the key factor of deterioration of drag control performance \cite{hammond1998observed}. 
% In addition, all the  Reynolds stresses are suppressed above the  virtual wall for the drag reduction cases, especially for the optimal control cases. 
% Compared to the uncontrolled case, the peak location moves away from the wall under control, which is consistent with the observation in the strictly incompressible cases \citep{choi1994active,deng2014strengthened} as well as for other control methods \citep{hurst2014effect,yao2019pof}.

 \subsection{Skin friction decomposition}
 The mean skin-friction $C_f$ can be decomposed into different physics-informed contributions based on the mean and statistical turbulence quantities across the wall layer.
 For example, the FIK (Fukagata, Iwamoto, Kasagi)
identity \citep{fukagata2002contribution,rastegari2015mechanism}  links $C_f$ with the total Reynolds shear stress $\langle \overline {u'v'}\rangle$, which is given as 
 \begin{eqnarray}\label{eq:FIK1}
C_f&=&\underbrace{\frac{6}{Re_b}}_{C^L_{f,F}}+\underbrace{\frac{3}{U^2_b}\int^1_{-1} y\langle\overline{{u'}{v'}}\rangle \mathrm{d}y}_{C^T_{T,F}}.
\end{eqnarray}

The drag coefficient $C_f$ thus has a laminar part $C^L_{f,F}=6/{Re_b}$, which is the drag  for laminar flow with the same flow rate, and a turbulent part $C^T_{f,F}$, which is represented by the weighted integration of the total Reynolds shear stress (RSS) $\langle\overline{{u'}{v'}}\rangle$. 
As the latter is composed of a coherent  $\langle{\tilde{u}\tilde{v}}\rangle$
and a random  part $\langle{\overline{{u''}{v''}}}\rangle$, Eq. \ref{eq:FIK1}  can be further decomposed as
\begin{eqnarray}\label{eq:FIK2}
C_f&=&\underbrace{\frac{6}{Re_b}}_{C^L_{f,F}}+\underbrace{\frac{3}{U^2_b}\int^1_{-1} y\langle{\tilde{u}\tilde{v}}\rangle \mathrm{d}y}_{C^C_{f,F}}
+\underbrace{\frac{3}{U^2_b}\int^1_{-1} y\langle\overline{{u''}{v''}}\rangle \mathrm{d}y}_{C^R_{f,F}}.
\end{eqnarray}

\begin{figure*}
\centering
       \subfloat[]{
      \includegraphics[width=0.46\textwidth]{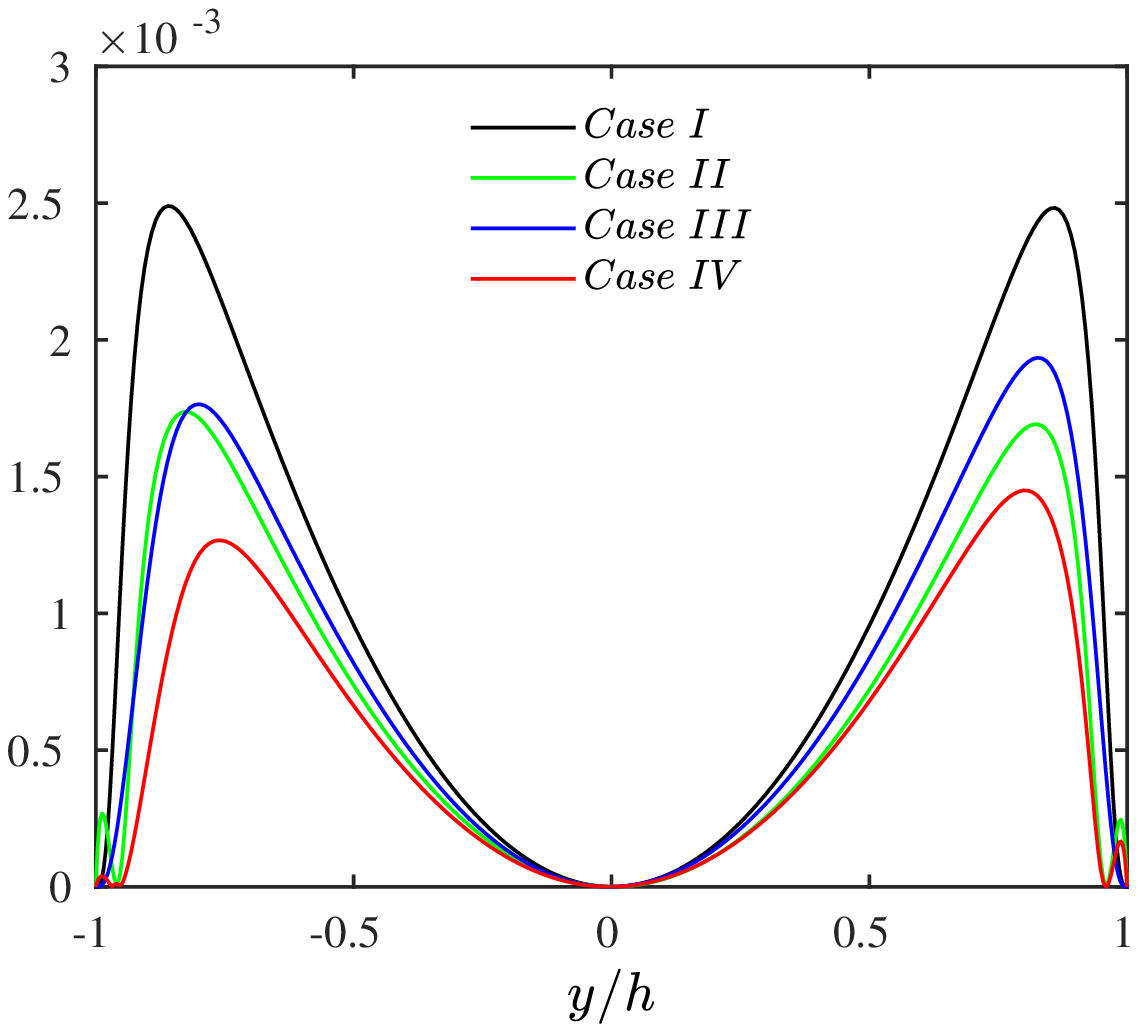}}
      \subfloat[]{
      \includegraphics[width=0.49\textwidth]{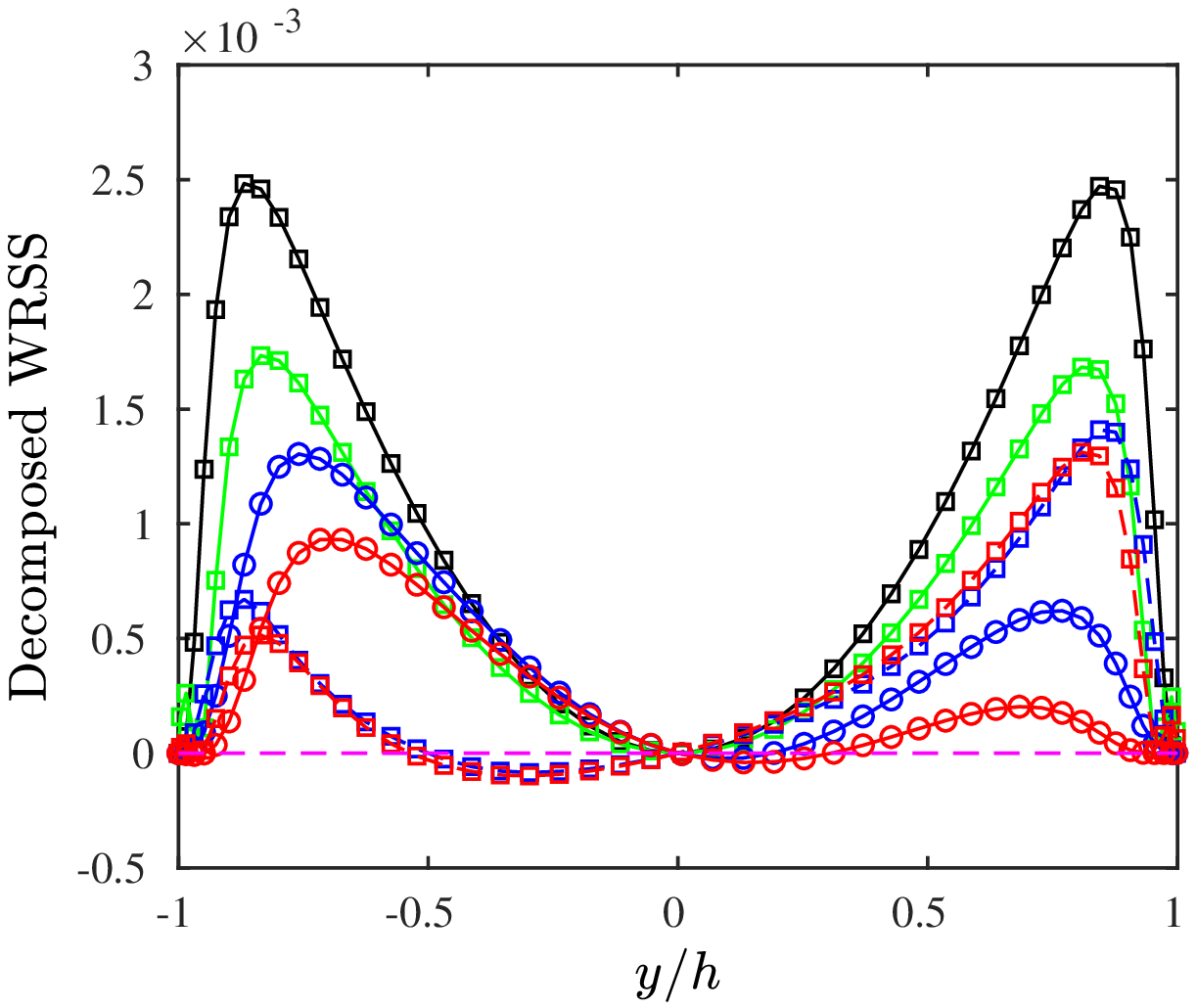}}
\caption{ (a) Total  and (b) decomposed weighted Reynolds shear stress (WRSS) distributions for different control cases.
 The symbols  \protect\marksymbol{o}{black} and \protect\marksymbol{square}{black} in (b) represent the coherent and random weighted Reynolds shear stress, respectively. }
\label{fig:WRSS}
\end{figure*}

Since $Re_b$ is held constant for all cases,
only changes in the coherent $C^C_{f,F}$ and random $C^R_{f,F}$ contributions affect the drag.
Figure \ref{fig:WRSS}(a) shows the profiles of the total weighted Reynolds shear stress (WRSS) $-(1-y)\langle{\overline{{u'}{v'}}}\rangle$,
and figure \ref{fig:WRSS}(b) shows the
corresponding coherent part $-(1-y)\langle{\tilde{u}\tilde{v}}\rangle$ (denoted as  C-WRSS)
and the random  part $-(1-y)\langle{\overline{{u''}{v''}}}\rangle$ (denoted as R-WRSS).
Following Eq. \eqref{eq:FIK1}, for all controlled cases, the drag reduction is mainly due to suppression  of  Reynolds shear stress in the near-wall region.
For Case II, the weighted Reynolds shear stress (WRSS) becomes almost zero  at the location of the virtual wall. 
For Case III, although SOJF is applied only near the bottom wall, the WRSS is also notably decreased near the top wall. 
Consistent with a larger $\mathcal{R}$, WRSS for Case IV is significantly suppressed   compared to the  uncontrolled case,  particularly in the bottom half of the channel. 
Furthermore, from figure \ref{fig:WRSS}(b),  C-WRSS dominates R-WRSS to become the main contributor to  WRSS in the bottom half channel, but the trend reverses in the top half channel. 
Interestingly, the main difference between Cases III and IV is   on C-WRSS.
It suggests that when compared to the SOJF, the better performance of CDC   is mainly due to smaller Reynolds shear stress associated with the weakened large-scale control swirls.

% the random turbulent Reynolds shear stress is significantly suppressed,
% and the coherent part .

\begin{figure*}
\centering
\includegraphics[width=0.9\textwidth]{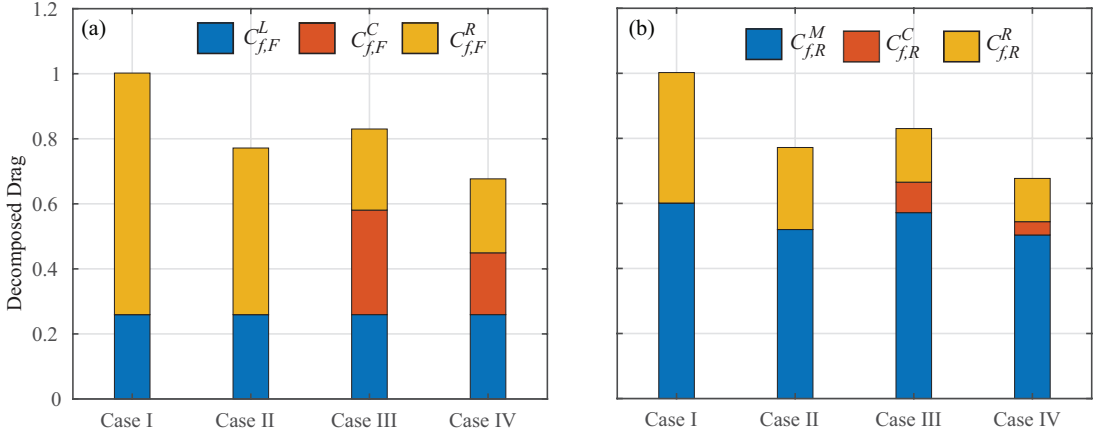}
%  \subfloat[]{
%       \includegraphics[width=0.48\textwidth]{FIK_d}}
%       \subfloat[]{
%       \includegraphics[width=0.48\textwidth]{FIK_R}}
\caption{The skin friction decomposition based on (a) FIK and (b) RD identities   for different cases. }
\label{fig:FIK}
\end{figure*}

Figure \ref{fig:FIK}(a) shows the decomposed total skin friction (normalized by $C_{f,0}$) based on the FIK identity. 
For the uncontrolled case (Case I), the laminar ($C^L_{f,F}$) and turbulent ($C^T_{f,F}$) parts contribute 25.8 \% and  74.2 \%  to the total $C_{f,0}$, respectively. For Case II, $C^T_{f,F}$ decreases to approximately  51.1\% of $C_{f,0}$. 
For Case III,  the random part $C^R_{f,F}$ contributes only about 24.9\% of $C_{f,0}$, while the coherent part $C^C_{f,F}$ becomes the dominant contribution,  about $32.1\%$.
When compared to Case III, both $C^R_{f,F}$  and $C^C_{f,F}$ are further decreased for Case IV,  contributing about  $19.0\%$ and $22.7\%$, respectively.

The FIK identity are often criticized due to a lack of  physical interpretation for the linear weight of Reynolds shear stress.
\citeauthor{renard2016theoretical} \cite{renard2016theoretical} proposed
an alternative mean skin friction decomposition, referred to as the RD identity. 
It is   derived from the mean streamwise kinetic-energy equation in an absolute reference frame and  characterizes the power of skin-friction as an energy transfer from the wall to the fluid by means of dissipation by molecular viscosity and turbulent production. 
In the case of turbulent channel flow, the RD identity can be written as 
\begin{widetext}
\begin{eqnarray}\label{eqn:RD}
 C_f&=&\underbrace{\frac{1}{U^3_b}\int^{1}_{-1}\nu\left(\frac{\partial  U}{\partial y}\right)^2\mathrm{d}y}_{C^M_{f,R}} +\underbrace{\frac{1}{U^3_b}\int^1_{-1} \langle\overline{-{u'}{v'}}\rangle\frac{\partial U}{\partial y}\mathrm{d}y}_{C^T_{f,R}}, \label{eqn:RD1}\\
 &=&\underbrace{\frac{1}{U^3_b}\int^{1}_{-1}\nu\left(\frac{\partial  U}{\partial y}\right)^2\mathrm{d}y}_{C^M_{f,R}} +\underbrace{\frac{1}{U^3_b}\int^1_{-1} \langle{-\tilde{u}\tilde{v}}\rangle\frac{\partial U}{\partial y}\mathrm{d}y}_{C^C_{f,R}}+\underbrace{\frac{1}{U^3_b}\int^1_{-1} \langle\overline{-{u''}{v''}}\rangle\frac{\partial U}{\partial y}\mathrm{d}y}_{C^R_{f,R}}, \label{eqn:RD2}
\end{eqnarray}
\end{widetext}
where $C^M_{f,R}$ represents the contribution from direct molecular viscous dissipation and $C^T_{f,R}$ characterizes the contribution associated with the turbulent-kinematic-energy (TKE) production $\langle -u'v' \rangle(\partial U/\partial y)$. Similarly, the turbulent contribution can be further separated into coherent ($C^C_{f,R}$) and random ($C^R_{f,R}$) parts, which are, respectively, associated with coherent and random Reynolds shear stresses. 
% \textcolor{blue}{remove: The RD identity overcomes some of the drawbacks of the FIK identity and provides a better physical interpretation of the skin-friction drag generation mechanism.} 

Figure \ref{fig:FIK}(b) shows the decomposed total skin friction (normalized by $C_{f,0}$) based on RD identity.
For the uncontrolled case (Case I),  $C^M_{f,R}$ overwhelm  $C^T_{f,R}$ at this low $Re$, contributing  59.9 \%  of $C_{f,0}$. 
Under OC, both  $C^M_{f,R}$ and  $C^T_{f,R}$ components are decreased to 51.8\% and 25.2\%, respectively. 
For case III, although the $C^T_{f,R}$ is almost the same as  Case II, the random contribution $C^R_{f,R}$ is only 16.5\% -- the rest is due to the coherent component $C^C_{f,R}$ that associated  with production of the large-scale control swirls. 
As discussed in Ref. \cite{yao2018drag}, effective drag reduction by SOJF results from a compensative effect of suppressing
random turbulence  countered by enhancing the coherent part induced by the forcing. 
For Case IV, $C^M_{f,R}$  is similar  to  Case II, but  $C^R_{f,R}$ is notably decreased. 
Compared to Case III, all three contributions are decreased, especially for $C^C_{C,R}$ due to smaller control amplitude.
It suggests that, with the aid of OC, SOJF becomes more effective due to the decreased counter effect caused by the large-scale coherent swirls. 

% For Case IV, although $C^M_{f,R}$ is almost the same as the uncontrolled case, the random contribution $C^R_{f,R}$ is significantly decreased. 
% In addition, the coherent part $C^C_{f,R}$ is much smaller than that of case III, and only  contributes about 4.2\% of $C_{f,0}$.

\subsection {Energy flux analysis} \label{sec:energ}

\begin{figure*}
\begin{centering}
{\includegraphics[width=0.8\textwidth]{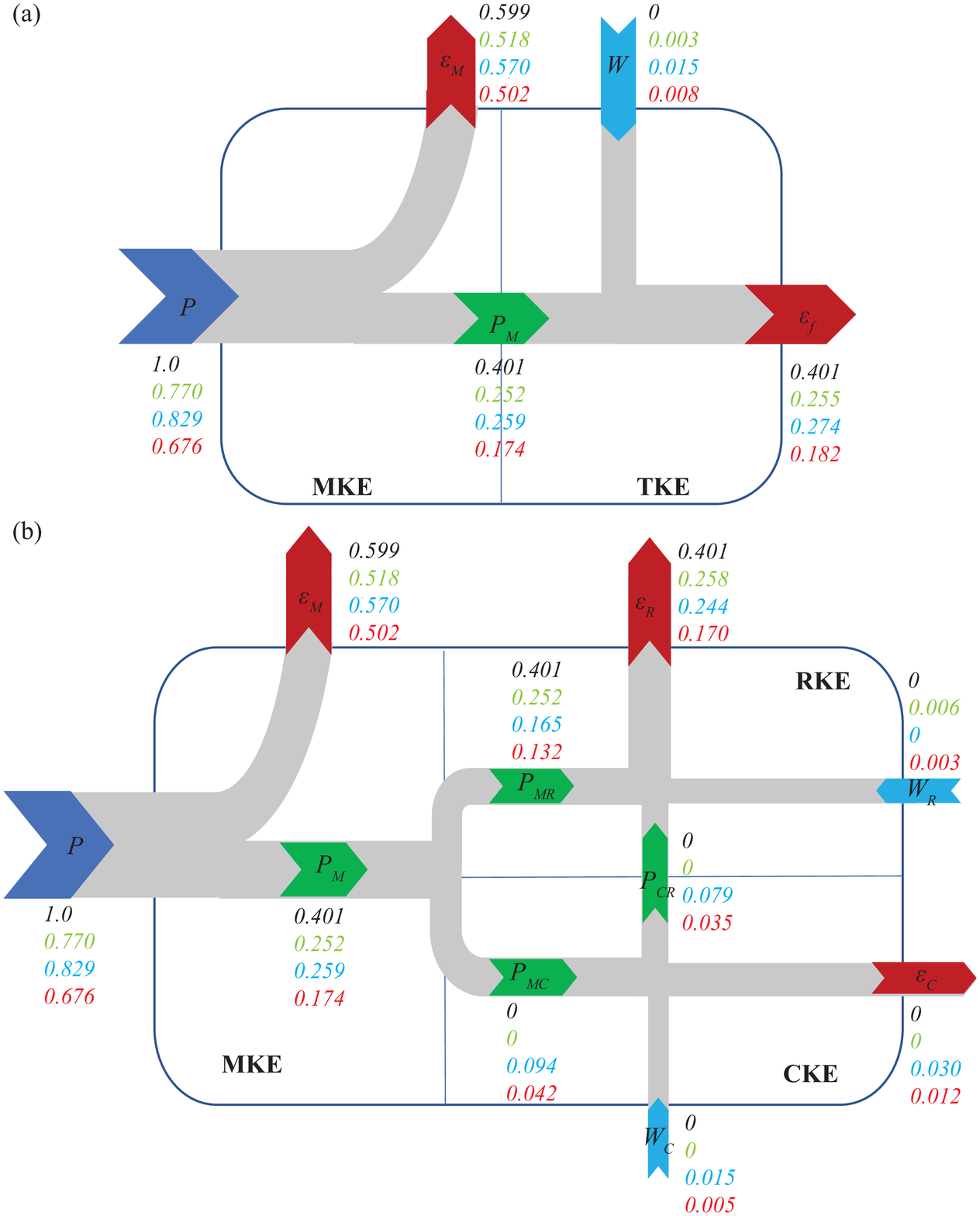}}
% {\includegraphics[width=0.48\textwidth]{Eb1c}}\\    %bud
\end{centering}
\caption{Energy flux box  by (a) Reynolds decomposition and (b) triple decomposition. Abbreviations MKE, TKE, CKE and RKE represent mean, turbulent, coherent, and random kinetic energies, respectively. 
All the budget terms have been normalized by the pumping power of the uncontrolled case, and their numerical values for Cases I to IV are presented  in order. }\label{fig:EnergBox}
\end{figure*}

% from top to low lines

For a flow system at statistic equilibrium, the rate at which energy enters the system (e.g., pumping power $P$ and control power input $W$) must equal the rate at which energy is dissipated (i.e., the viscous dissipation $\epsilon$). 
When employing the Reynolds decomposition, the kinetic energy of the flow can be separated into the kinetic energy of the mean  (MKE) and  turbulent fields (TKE).
In this section, we examine how the
energy transfer between the mean flow and the turbulent fluctuations is affected by control.

% In Chen, Yao \& Hussain \cite{chen2021}, t is expressed via the volume integrated dissipation and the external energy input, i.e.
%  \begin{equation}\label{eq:Cf:energy3}
%  \frac{C_f}{2}=\int^{1}_0 \frac{\epsilon_{_M}}{\epsilon_*}dy'+\int^{1}_0\frac{{\epsilon}_{_f}}{\epsilon_*}dy'
%  -\frac{W}{\epsilon_*}.
%  \end{equation}
%  Here,  $\epsilon_*=u^3_{b}/h$; $\epsilon_{_M}=\nu \partial_y \overline{u} \partial_y \overline{u}$ is the mean flow dissipation and $ {\epsilon}_{_f}=\nu\overline{\partial_j u'_i\partial_j u'_i}$ is turbulent dissipation. 
%  From Eq.\ref{eq:Cf:energy3}, the energy flux from mean flow to turbulent fluctuations presents another way to understand the drag reduction and net power saving mechanisms.
 
As suggested by \citeauthor{ricco2012changes}  \cite{ricco2012changes}, the energy flux can be compactly described  using the so-called  energy box, sketched in Figure \ref{fig:EnergBox}a. 
The pumping power is merely associated with the mean flow, while the energy dissipation is split into parts corresponding to the mean ($\epsilon_M$) and turbulent   ($\epsilon_f$) fields.  In addition, the energy transfer  from the mean to the fluctuating field is  embodied by the production  of turbulent kinetic energy $P_M$.
$P_M$ acts as a sink for MKE, but as a
source for TKE, which is equal to the sum of $\epsilon_f$ and $W$.
Hence, in contrast to RD identity,  \citeauthor{chen2021} \cite{chen2021}  provided an alternative expression for $C_f$ in terms of dissipation and control power input, i.e., $C_f=(\epsilon_M+\epsilon_f-W)/U^3_b$. 
Similar to that shown in Fig. \ref{fig:FIK}b, 
for the uncontrolled case,  59.9\% and 40.1\% of the energy that enters  the system (through pumping power)  are dissipated by the mean and turbulent fields, respectively. 
With control, the pumping power is decreased and the amount of decrease  corresponds to the rate of drag reduction $\mathcal{R}$ presented in Sec. \ref{sec:drag}.
Furthermore, the change of total dissipation ($\epsilon_M+\epsilon_f$) is the rate of net power saving $\mathcal{N}$ \cite{chen2021}.

For all the control cases, the dissipation due to the mean flow fields $\epsilon_M$ does not change much, which contributes slightly above 50\% of $P_0$. 
As a reference, the dissipation due to the laminar mean flow is about one-fourth of $P_0$ \cite{gatti2018global}.
Unlike $\epsilon_M$, the dissipation due to the turbulent field  $\epsilon_f$ vastly decreases under control -- suggesting significant suppression of near-wall random streamwise vortices. 
It is of particular interest to look at Case IV, where the control power input $W(=0.8\%)$ is in between Case II ($W=0.3\%$) and Case III($W=1.5\%$), but $\epsilon_f$ is much smaller than these two cases. 

\red{To better understand how control flow affects the energy transfer, \citeauthor{gatti2018global} \cite{gatti2018global}  developed an elaborated energy flux analysis by separating the mean flow into a laminar component and a deviation from it. }
Although it provides some further insight on the effect of control on  energy flux, it  lacks  a physical basis to separate the mean flow into laminar mean and the remaining parts.
Differently,  \citeauthor{chen2021} \cite{chen2021} introduced a triple-decomposition (eq. \ref{eq:trp}) into the energy budget analysis, which enables us to examine how controls affect the energy transfer among the mean, coherent, and turbulent fields.
The corresponding energy box is shown in Fig. \ref{fig:EnergBox}b.
For flow under control, the control power can be injected either through random velocity field $W_{_R}$ (e.g., for OC) or through coherent velocity field $W_{_C}$ (e.g., for SOJF).
Meanwhile, production converts mean flow energy to both coherent ($P_{MC}$) and random ($P_{MR}$) velocity fields.
There is an additional  production ($P_{CR}$)  associated with the energy transfer from coherent to the random velocity field. 
Note that similar energy budget analysis has recently been employed by \citeauthor{zonta2021} \cite{zonta2021} to understand drag reduction mechanisms of lubricated turbulent channel flows.

As the coherent part is absent for Cases I and II, the analysis is equivalent to that 
based on Reynolds decomposition.
\redsec{For both Cases III and IV, the dissipation due to coherent flow field  $\epsilon_C(=\langle\overline{\nu \partial_i\tilde{u}_j\partial_i\tilde{u}_j}\rangle)$
is rather small.} For Case III, although the majority of the turbulent production $P_M$ directly goes to the random turbulent
field, around 40\% of $P_M$ is injected into the coherent field, most of which, together with the coherent control power input $W_C$, is eventually transferred to the random fields and is dissipated as  $\epsilon_R(=\langle\overline{\nu \partial_iu''_j\partial_iu''_j}\rangle)$.
For Case IV, due to a smaller forcing amplitude $A^+_s$, the large-scale control swirls generated are weaker than Case III. As a consequence, the production and dissipation associated with that are significantly decreased. 
In addition, although the detecting plane $y^+_d$ is kept the same, the random  control power input $W_R$ is decreased compared to Case II. This is not unexpected as  the large-scale swirls weaken the low-speed streaks and  suppress the random turbulent fluctuations.

\section{Discussion and Conclusion}\label{sec:con}

Composite drag control (CDC) based on the  opposition control (OC) and spanwise opposed wall-jet forcing (SOJF) methods is investigated via direct numerical simulation of the incompressible Navier–Stokes equations in a turbulent channel flow.
A limited parameter search at $Re_\tau=180$ shows that a maximum $\mathcal{R}$ of  $32\%$ can be obtained with $y^+_d=15$ and $A^+_s=0.0067$, which is much higher than that  can be achieved by each individual method.
Furthermore, due to the small power input required, significant net power saving can be achieved. 
Flow analysis, including skin friction decomposition and flow visualization,  reveals that CDC can take advantage of both OC and SOJF methods. 
In particular, compared to OC, it is more effective in suppressing the  random turbulence  due to the presence of large-scale coherent swirls generated by SOJF.
In addition, compared to SOJF, it requires  weaker large-scale coherent swirls due to the role  of OC in suppressing random turbulence -- hence decreasing the drag contribution associated with these swirls.

The imposition of SOJF improves the coherence between  the measurable wall quantities and the velocity fluctuations in the flow. 
This also indicates that this control strategy could also yield better performance for more practical wall-quantity-based drag control methods, such as the sub-optimal control and those based on machine learning. 
For most of the drag control methods, the performance decreases as $Re$ increases, mainly due to the increased contributions of large-scale and very-large-scale structures \cite{marusic2010predictive}. 
As OC and SOJF are designed to suppress small- and large-scale structures, respectively, this CDC strategy might be employed as one possible way to overcome the deterioration of drag reduction performance at high $Re$.
All these questions deserve further investigation.

\begin{acknowledgments}
Computational resources provided by  Texas Tech University HPCC, TACC Lonestar are acknowledged.
We appreciate Lee and Moser for providing us their DNS code, which is used for this study. X.C. appreciates the funding support by the National Natural Science Foundation of China No. 12072012, 11721202, 91952302.

\end{acknowledgments}

%% Harvard

\bibliography{jfm-instructions2}

\begin{thebibliography}{58}%
\makeatletter
\providecommand \@ifxundefined [1]{%
 \@ifx{#1\undefined}
}%
\providecommand \@ifnum [1]{%
 \ifnum #1\expandafter \@firstoftwo
 \else \expandafter \@secondoftwo
 \fi
}%
\providecommand \@ifx [1]{%
 \ifx #1\expandafter \@firstoftwo
 \else \expandafter \@secondoftwo
 \fi
}%
\providecommand \natexlab [1]{#1}%
\providecommand \enquote  [1]{``#1''}%
\providecommand \bibnamefont  [1]{#1}%
\providecommand \bibfnamefont [1]{#1}%
\providecommand \citenamefont [1]{#1}%
\providecommand \href@noop [0]{\@secondoftwo}%
\providecommand \href [0]{\begingroup \@sanitize@url \@href}%
\providecommand \@href[1]{\@@startlink{#1}\@@href}%
\providecommand \@@href[1]{\endgroup#1\@@endlink}%
\providecommand \@sanitize@url [0]{\catcode `\\12\catcode `\$12\catcode
  `\&12\catcode `\#12\catcode `\^12\catcode `\_12\catcode `\%12\relax}%
\providecommand \@@startlink[1]{}%
\providecommand \@@endlink[0]{}%
\providecommand \url  [0]{\begingroup\@sanitize@url \@url }%
\providecommand \@url [1]{\endgroup\@href {#1}{\urlprefix }}%
\providecommand \urlprefix  [0]{URL }%
\providecommand \Eprint [0]{\href }%
\providecommand \doibase [0]{https://doi.org/}%
\providecommand \selectlanguage [0]{\@gobble}%
\providecommand \bibinfo  [0]{\@secondoftwo}%
\providecommand \bibfield  [0]{\@secondoftwo}%
\providecommand \translation [1]{[#1]}%
\providecommand \BibitemOpen [0]{}%
\providecommand \bibitemStop [0]{}%
\providecommand \bibitemNoStop [0]{.\EOS\space}%
\providecommand \EOS [0]{\spacefactor3000\relax}%
\providecommand \BibitemShut  [1]{\csname bibitem#1\endcsname}%
\let\auto@bib@innerbib\@empty
%</preamble>
\bibitem [{\citenamefont {Waleffe}(1997)}]{waleffe1997self}%
  \BibitemOpen
  \bibfield  {author} {\bibinfo {author} {\bibfnamefont {F.}~\bibnamefont
  {Waleffe}},\ }\bibfield  {title} {\bibinfo {title} {On a self-sustaining
  process in shear flows},\ }\href@noop {} {\bibfield  {journal} {\bibinfo
  {journal} {Physics of Fluids}\ }\textbf {\bibinfo {volume} {9}},\ \bibinfo
  {pages} {883} (\bibinfo {year} {1997})}\BibitemShut {NoStop}%
\bibitem [{\citenamefont {Jim{\'e}nez}\ and\ \citenamefont
  {Pinelli}(1999)}]{jimenez1999autonomous}%
  \BibitemOpen
  \bibfield  {author} {\bibinfo {author} {\bibfnamefont {J.}~\bibnamefont
  {Jim{\'e}nez}}\ and\ \bibinfo {author} {\bibfnamefont {A.}~\bibnamefont
  {Pinelli}},\ }\bibfield  {title} {\bibinfo {title} {The autonomous cycle of
  near-wall turbulence},\ }\href@noop {} {\bibfield  {journal} {\bibinfo
  {journal} {Journal of Fluid Mechanics}\ }\textbf {\bibinfo {volume} {389}},\
  \bibinfo {pages} {335} (\bibinfo {year} {1999})}\BibitemShut {NoStop}%
\bibitem [{\citenamefont {Schoppa}\ and\ \citenamefont
  {Hussain}(2002)}]{schoppa2002coherent}%
  \BibitemOpen
  \bibfield  {author} {\bibinfo {author} {\bibfnamefont {W.}~\bibnamefont
  {Schoppa}}\ and\ \bibinfo {author} {\bibfnamefont {F.}~\bibnamefont
  {Hussain}},\ }\bibfield  {title} {\bibinfo {title} {Coherent structure
  generation in near-wall turbulence},\ }\href@noop {} {\bibfield  {journal}
  {\bibinfo  {journal} {Journal of fluid Mechanics}\ }\textbf {\bibinfo
  {volume} {453}},\ \bibinfo {pages} {57} (\bibinfo {year} {2002})}\BibitemShut
  {NoStop}%
\bibitem [{\citenamefont {Gad-el Hak}(2007)}]{gad2007flow}%
  \BibitemOpen
  \bibfield  {author} {\bibinfo {author} {\bibfnamefont {M.}~\bibnamefont
  {Gad-el Hak}},\ }\href@noop {} {\emph {\bibinfo {title} {Flow control:
  passive, active, and reactive flow management}}}\ (\bibinfo  {publisher}
  {Cambridge University Press},\ \bibinfo {year} {2007})\BibitemShut {NoStop}%
\bibitem [{\citenamefont {Bechert}\ \emph {et~al.}(1997)\citenamefont
  {Bechert}, \citenamefont {Bruse}, \citenamefont {Hage}, \citenamefont
  {Van~der Hoeven},\ and\ \citenamefont {Hoppe}}]{bechert1997experiments}%
  \BibitemOpen
  \bibfield  {author} {\bibinfo {author} {\bibfnamefont {D.}~\bibnamefont
  {Bechert}}, \bibinfo {author} {\bibfnamefont {M.}~\bibnamefont {Bruse}},
  \bibinfo {author} {\bibfnamefont {W.~v.}\ \bibnamefont {Hage}}, \bibinfo
  {author} {\bibfnamefont {J.~T.}\ \bibnamefont {Van~der Hoeven}},\ and\
  \bibinfo {author} {\bibfnamefont {G.}~\bibnamefont {Hoppe}},\ }\bibfield
  {title} {\bibinfo {title} {Experiments on drag-reducing surfaces and their
  optimization with an adjustable geometry},\ }\href@noop {} {\bibfield
  {journal} {\bibinfo  {journal} {Journal of fluid mechanics}\ }\textbf
  {\bibinfo {volume} {338}},\ \bibinfo {pages} {59} (\bibinfo {year}
  {1997})}\BibitemShut {NoStop}%
\bibitem [{\citenamefont {Rothstein}(2010)}]{rothstein2010slip}%
  \BibitemOpen
  \bibfield  {author} {\bibinfo {author} {\bibfnamefont {J.~P.}\ \bibnamefont
  {Rothstein}},\ }\bibfield  {title} {\bibinfo {title} {Slip on
  superhydrophobic surfaces},\ }\href@noop {} {\bibfield  {journal} {\bibinfo
  {journal} {Annual Review of Fluid Mechanics}\ }\textbf {\bibinfo {volume}
  {42}},\ \bibinfo {pages} {89} (\bibinfo {year} {2010})}\BibitemShut {NoStop}%
\bibitem [{\citenamefont {Park}\ \emph {et~al.}(2014)\citenamefont {Park},
  \citenamefont {Sun} \emph {et~al.}}]{park2014superhydrophobic}%
  \BibitemOpen
  \bibfield  {author} {\bibinfo {author} {\bibfnamefont {H.}~\bibnamefont
  {Park}}, \bibinfo {author} {\bibfnamefont {G.}~\bibnamefont {Sun}}, \emph
  {et~al.},\ }\bibfield  {title} {\bibinfo {title} {Superhydrophobic turbulent
  drag reduction as a function of surface grating parameters},\ }\href@noop {}
  {\bibfield  {journal} {\bibinfo  {journal} {Journal of Fluid Mechanics}\
  }\textbf {\bibinfo {volume} {747}},\ \bibinfo {pages} {722} (\bibinfo {year}
  {2014})}\BibitemShut {NoStop}%
\bibitem [{\citenamefont {Checco}\ \emph {et~al.}(2014)\citenamefont {Checco},
  \citenamefont {Ocko}, \citenamefont {Rahman}, \citenamefont {Black},
  \citenamefont {Tasinkevych}, \citenamefont {Giacomello},\ and\ \citenamefont
  {Dietrich}}]{checco2014collapse}%
  \BibitemOpen
  \bibfield  {author} {\bibinfo {author} {\bibfnamefont {A.}~\bibnamefont
  {Checco}}, \bibinfo {author} {\bibfnamefont {B.~M.}\ \bibnamefont {Ocko}},
  \bibinfo {author} {\bibfnamefont {A.}~\bibnamefont {Rahman}}, \bibinfo
  {author} {\bibfnamefont {C.~T.}\ \bibnamefont {Black}}, \bibinfo {author}
  {\bibfnamefont {M.}~\bibnamefont {Tasinkevych}}, \bibinfo {author}
  {\bibfnamefont {A.}~\bibnamefont {Giacomello}},\ and\ \bibinfo {author}
  {\bibfnamefont {S.}~\bibnamefont {Dietrich}},\ }\bibfield  {title} {\bibinfo
  {title} {Collapse and reversibility of the superhydrophobic state on
  nanotextured surfaces},\ }\href@noop {} {\bibfield  {journal} {\bibinfo
  {journal} {Physical Review Letters}\ }\textbf {\bibinfo {volume} {112}},\
  \bibinfo {pages} {216101} (\bibinfo {year} {2014})}\BibitemShut {NoStop}%
\bibitem [{\citenamefont {Rastegari}\ and\ \citenamefont
  {Akhavan}(2018)}]{rastegari2018common}%
  \BibitemOpen
  \bibfield  {author} {\bibinfo {author} {\bibfnamefont {A.}~\bibnamefont
  {Rastegari}}\ and\ \bibinfo {author} {\bibfnamefont {R.}~\bibnamefont
  {Akhavan}},\ }\bibfield  {title} {\bibinfo {title} {The common mechanism of
  turbulent skin-friction drag reduction with superhydrophobic longitudinal
  microgrooves and riblets},\ }\href@noop {} {\bibfield  {journal} {\bibinfo
  {journal} {Journal of Fluid Mechanics}\ }\textbf {\bibinfo {volume} {838}},\
  \bibinfo {pages} {68} (\bibinfo {year} {2018})}\BibitemShut {NoStop}%
\bibitem [{\citenamefont {Sumitani}\ and\ \citenamefont
  {Kasagi}(1995)}]{sumitani1995direct}%
  \BibitemOpen
  \bibfield  {author} {\bibinfo {author} {\bibfnamefont {Y.}~\bibnamefont
  {Sumitani}}\ and\ \bibinfo {author} {\bibfnamefont {N.}~\bibnamefont
  {Kasagi}},\ }\bibfield  {title} {\bibinfo {title} {Direct numerical
  simulation of turbulent transport with uniform wall injection and suction},\
  }\href@noop {} {\bibfield  {journal} {\bibinfo  {journal} {AIAA journal}\
  }\textbf {\bibinfo {volume} {33}},\ \bibinfo {pages} {1220} (\bibinfo {year}
  {1995})}\BibitemShut {NoStop}%
\bibitem [{\citenamefont {Kametani}\ and\ \citenamefont
  {Fukagata}(2011)}]{kametani2011direct}%
  \BibitemOpen
  \bibfield  {author} {\bibinfo {author} {\bibfnamefont {Y.}~\bibnamefont
  {Kametani}}\ and\ \bibinfo {author} {\bibfnamefont {K.}~\bibnamefont
  {Fukagata}},\ }\bibfield  {title} {\bibinfo {title} {Direct numerical
  simulation of spatially developing turbulent boundary layers with uniform
  blowing or suction},\ }\href@noop {} {\bibfield  {journal} {\bibinfo
  {journal} {Journal of Fluid Mechanics}\ }\textbf {\bibinfo {volume} {681}},\
  \bibinfo {pages} {154} (\bibinfo {year} {2011})}\BibitemShut {NoStop}%
\bibitem [{\citenamefont {Jung}\ \emph {et~al.}(1992)\citenamefont {Jung},
  \citenamefont {Mangiavacchi},\ and\ \citenamefont
  {Akhavan}}]{jung1992suppression}%
  \BibitemOpen
  \bibfield  {author} {\bibinfo {author} {\bibfnamefont {W.}~\bibnamefont
  {Jung}}, \bibinfo {author} {\bibfnamefont {N.}~\bibnamefont {Mangiavacchi}},\
  and\ \bibinfo {author} {\bibfnamefont {R.}~\bibnamefont {Akhavan}},\
  }\bibfield  {title} {\bibinfo {title} {Suppression of turbulence in
  wall-bounded flows by high-frequency spanwise oscillations},\ }\href@noop {}
  {\bibfield  {journal} {\bibinfo  {journal} {Physics of Fluids A: Fluid
  Dynamics}\ }\textbf {\bibinfo {volume} {4}},\ \bibinfo {pages} {1605}
  (\bibinfo {year} {1992})}\BibitemShut {NoStop}%
\bibitem [{\citenamefont {Choi}\ \emph {et~al.}(2002)\citenamefont {Choi},
  \citenamefont {Xu},\ and\ \citenamefont {Sung}}]{choi2002drag}%
  \BibitemOpen
  \bibfield  {author} {\bibinfo {author} {\bibfnamefont {J.-I.}\ \bibnamefont
  {Choi}}, \bibinfo {author} {\bibfnamefont {C.-X.}\ \bibnamefont {Xu}},\ and\
  \bibinfo {author} {\bibfnamefont {H.~J.}\ \bibnamefont {Sung}},\ }\bibfield
  {title} {\bibinfo {title} {Drag reduction by spanwise wall oscillation in
  wall-bounded turbulent flows},\ }\href@noop {} {\bibfield  {journal}
  {\bibinfo  {journal} {AIAA journal}\ }\textbf {\bibinfo {volume} {40}},\
  \bibinfo {pages} {842} (\bibinfo {year} {2002})}\BibitemShut {NoStop}%
\bibitem [{\citenamefont {Quadrio}\ \emph {et~al.}(2009)\citenamefont
  {Quadrio}, \citenamefont {Ricco},\ and\ \citenamefont
  {Viotti}}]{quadrio2009streamwise}%
  \BibitemOpen
  \bibfield  {author} {\bibinfo {author} {\bibfnamefont {M.}~\bibnamefont
  {Quadrio}}, \bibinfo {author} {\bibfnamefont {P.}~\bibnamefont {Ricco}},\
  and\ \bibinfo {author} {\bibfnamefont {C.}~\bibnamefont {Viotti}},\
  }\bibfield  {title} {\bibinfo {title} {Streamwise-travelling waves of
  spanwise wall velocity for turbulent drag reduction},\ }\href@noop {}
  {\bibfield  {journal} {\bibinfo  {journal} {Journal of Fluid Mechanics}\
  }\textbf {\bibinfo {volume} {627}},\ \bibinfo {pages} {161} (\bibinfo {year}
  {2009})}\BibitemShut {NoStop}%
\bibitem [{\citenamefont {Quadrio}(2011)}]{Quadrio2011}%
  \BibitemOpen
  \bibfield  {author} {\bibinfo {author} {\bibfnamefont {M.}~\bibnamefont
  {Quadrio}},\ }\bibfield  {title} {\bibinfo {title} {Drag reduction in
  turbulent boundary layers by in-plane wall motion},\ }\href@noop {}
  {\bibfield  {journal} {\bibinfo  {journal} {Philosophical Transactions of the
  Royal Society A}\ ,\ \bibinfo {pages} {1428}} (\bibinfo {year}
  {2011})}\BibitemShut {NoStop}%
\bibitem [{\citenamefont {Agostini}\ \emph {et~al.}(2014)\citenamefont
  {Agostini}, \citenamefont {Touber},\ and\ \citenamefont
  {Leschziner}}]{Agostini2014}%
  \BibitemOpen
  \bibfield  {author} {\bibinfo {author} {\bibfnamefont {L.}~\bibnamefont
  {Agostini}}, \bibinfo {author} {\bibfnamefont {E.}~\bibnamefont {Touber}},\
  and\ \bibinfo {author} {\bibfnamefont {M.~A.}\ \bibnamefont {Leschziner}},\
  }\bibfield  {title} {\bibinfo {title} {Spanwise oscillatory wall motion in
  channel flow: drag-reduction mechanisms inferred from dns-predicted
  phase-wise property variations at ret=1000},\ }\href@noop {} {\bibfield
  {journal} {\bibinfo  {journal} {Journal of Fluid Mechanics}\ ,\ \bibinfo
  {pages} {606}} (\bibinfo {year} {2014})}\BibitemShut {NoStop}%
\bibitem [{\citenamefont {Yakeno}\ \emph {et~al.}(2014)\citenamefont {Yakeno},
  \citenamefont {Hasegawa},\ and\ \citenamefont
  {Kasagi}}]{yakeno2014modification}%
  \BibitemOpen
  \bibfield  {author} {\bibinfo {author} {\bibfnamefont {A.}~\bibnamefont
  {Yakeno}}, \bibinfo {author} {\bibfnamefont {Y.}~\bibnamefont {Hasegawa}},\
  and\ \bibinfo {author} {\bibfnamefont {N.}~\bibnamefont {Kasagi}},\
  }\bibfield  {title} {\bibinfo {title} {Modification of quasi-streamwise
  vortical structure in a drag-reduced turbulent channel flow with spanwise
  wall oscillation},\ }\href@noop {} {\bibfield  {journal} {\bibinfo  {journal}
  {Physics of Fluids}\ }\textbf {\bibinfo {volume} {26}},\ \bibinfo {pages}
  {085109} (\bibinfo {year} {2014})}\BibitemShut {NoStop}%
\bibitem [{\citenamefont {Choi}\ \emph {et~al.}(1994)\citenamefont {Choi},
  \citenamefont {Moin},\ and\ \citenamefont {Kim}}]{choi1994active}%
  \BibitemOpen
  \bibfield  {author} {\bibinfo {author} {\bibfnamefont {H.}~\bibnamefont
  {Choi}}, \bibinfo {author} {\bibfnamefont {P.}~\bibnamefont {Moin}},\ and\
  \bibinfo {author} {\bibfnamefont {J.}~\bibnamefont {Kim}},\ }\bibfield
  {title} {\bibinfo {title} {Active turbulence control for drag reduction in
  wall-bounded flows},\ }\href@noop {} {\bibfield  {journal} {\bibinfo
  {journal} {Journal of Fluid Mechanics}\ }\textbf {\bibinfo {volume} {262}},\
  \bibinfo {pages} {75} (\bibinfo {year} {1994})}\BibitemShut {NoStop}%
\bibitem [{\citenamefont {Kim}\ and\ \citenamefont
  {Bewley}(2007)}]{kim2007linear}%
  \BibitemOpen
  \bibfield  {author} {\bibinfo {author} {\bibfnamefont {J.}~\bibnamefont
  {Kim}}\ and\ \bibinfo {author} {\bibfnamefont {T.~R.}\ \bibnamefont
  {Bewley}},\ }\bibfield  {title} {\bibinfo {title} {A linear systems approach
  to flow control},\ }\href@noop {} {\bibfield  {journal} {\bibinfo  {journal}
  {Annu. Rev. Fluid Mech.}\ }\textbf {\bibinfo {volume} {39}},\ \bibinfo
  {pages} {383} (\bibinfo {year} {2007})}\BibitemShut {NoStop}%
\bibitem [{\citenamefont {Chung}\ and\ \citenamefont
  {Talha}(2011)}]{chung2011effectiveness}%
  \BibitemOpen
  \bibfield  {author} {\bibinfo {author} {\bibfnamefont {Y.~M.}\ \bibnamefont
  {Chung}}\ and\ \bibinfo {author} {\bibfnamefont {T.}~\bibnamefont {Talha}},\
  }\bibfield  {title} {\bibinfo {title} {Effectiveness of active flow control
  for turbulent skin friction drag reduction},\ }\href@noop {} {\bibfield
  {journal} {\bibinfo  {journal} {Physics of Fluids (1994-present)}\ }\textbf
  {\bibinfo {volume} {23}},\ \bibinfo {pages} {025102} (\bibinfo {year}
  {2011})}\BibitemShut {NoStop}%
\bibitem [{\citenamefont {Deng}\ and\ \citenamefont {Xu}(2012)}]{deng2012}%
  \BibitemOpen
  \bibfield  {author} {\bibinfo {author} {\bibfnamefont {B.-Q.}\ \bibnamefont
  {Deng}}\ and\ \bibinfo {author} {\bibfnamefont {C.-X.}\ \bibnamefont {Xu}},\
  }\bibfield  {title} {\bibinfo {title} {Influence of active control on
  stg-based generation of streamwise vortices in near-wall turbulence},\ }\href
  {https://doi.org/10.1017/jfm.2012.361} {\bibfield  {journal} {\bibinfo
  {journal} {Journal of Fluid Mechanics}\ }\textbf {\bibinfo {volume} {710}},\
  \bibinfo {pages} {234} (\bibinfo {year} {2012})}\BibitemShut {NoStop}%
\bibitem [{\citenamefont {Mamori}\ \emph {et~al.}(2014)\citenamefont {Mamori},
  \citenamefont {Iwamoto},\ and\ \citenamefont {Murata}}]{mamori2014effect}%
  \BibitemOpen
  \bibfield  {author} {\bibinfo {author} {\bibfnamefont {H.}~\bibnamefont
  {Mamori}}, \bibinfo {author} {\bibfnamefont {K.}~\bibnamefont {Iwamoto}},\
  and\ \bibinfo {author} {\bibfnamefont {A.}~\bibnamefont {Murata}},\
  }\bibfield  {title} {\bibinfo {title} {Effect of the parameters of traveling
  waves created by blowing and suction on the relaminarization phenomena in
  fully developed turbulent channel flow},\ }\href@noop {} {\bibfield
  {journal} {\bibinfo  {journal} {Physics of Fluids (1994-present)}\ }\textbf
  {\bibinfo {volume} {26}},\ \bibinfo {pages} {015101} (\bibinfo {year}
  {2014})}\BibitemShut {NoStop}%
\bibitem [{\citenamefont {Gatti}\ and\ \citenamefont
  {Quadrio}(2016)}]{gatti2016reynolds}%
  \BibitemOpen
  \bibfield  {author} {\bibinfo {author} {\bibfnamefont {D.}~\bibnamefont
  {Gatti}}\ and\ \bibinfo {author} {\bibfnamefont {M.}~\bibnamefont
  {Quadrio}},\ }\bibfield  {title} {\bibinfo {title} {Reynolds-number
  dependence of turbulent skin-friction drag reduction induced by spanwise
  forcing},\ }\href@noop {} {\bibfield  {journal} {\bibinfo  {journal} {Journal
  of Fluid Mechanics}\ }\textbf {\bibinfo {volume} {802}},\ \bibinfo {pages}
  {553} (\bibinfo {year} {2016})}\BibitemShut {NoStop}%
\bibitem [{\citenamefont {Yao}\ \emph {et~al.}(2019)\citenamefont {Yao},
  \citenamefont {Chen},\ and\ \citenamefont {Hussain}}]{yao2019reynolds}%
  \BibitemOpen
  \bibfield  {author} {\bibinfo {author} {\bibfnamefont {J.}~\bibnamefont
  {Yao}}, \bibinfo {author} {\bibfnamefont {X.}~\bibnamefont {Chen}},\ and\
  \bibinfo {author} {\bibfnamefont {F.}~\bibnamefont {Hussain}},\ }\bibfield
  {title} {\bibinfo {title} {Reynolds number effect on drag control via
  spanwise wall oscillation in turbulent channel flows},\ }\href@noop {}
  {\bibfield  {journal} {\bibinfo  {journal} {Physics of Fluids}\ }\textbf
  {\bibinfo {volume} {31}},\ \bibinfo {pages} {085108} (\bibinfo {year}
  {2019})}\BibitemShut {NoStop}%
\bibitem [{\citenamefont {Min}\ \emph {et~al.}(2006)\citenamefont {Min},
  \citenamefont {Kang}, \citenamefont {Speyer},\ and\ \citenamefont
  {Kim}}]{min2006sustained}%
  \BibitemOpen
  \bibfield  {author} {\bibinfo {author} {\bibfnamefont {T.}~\bibnamefont
  {Min}}, \bibinfo {author} {\bibfnamefont {S.~M.}\ \bibnamefont {Kang}},
  \bibinfo {author} {\bibfnamefont {J.~L.}\ \bibnamefont {Speyer}},\ and\
  \bibinfo {author} {\bibfnamefont {J.}~\bibnamefont {Kim}},\ }\bibfield
  {title} {\bibinfo {title} {Sustained sub-laminar drag in a fully developed
  channel flow},\ }\href@noop {} {\bibfield  {journal} {\bibinfo  {journal}
  {Journal of Fluid Mechanics}\ }\textbf {\bibinfo {volume} {558}},\ \bibinfo
  {pages} {309} (\bibinfo {year} {2006})}\BibitemShut {NoStop}%
\bibitem [{\citenamefont {Moarref}\ and\ \citenamefont
  {Jovanovi{\'c}}(2010)}]{moarref2010controlling}%
  \BibitemOpen
  \bibfield  {author} {\bibinfo {author} {\bibfnamefont {R.}~\bibnamefont
  {Moarref}}\ and\ \bibinfo {author} {\bibfnamefont {M.~R.}\ \bibnamefont
  {Jovanovi{\'c}}},\ }\bibfield  {title} {\bibinfo {title} {Controlling the
  onset of turbulence by streamwise traveling waves. part 1: Receptivity
  analysis},\ }\href@noop {} {\bibfield  {journal} {\bibinfo  {journal} {arXiv
  preprint arXiv:1006.4594}\ } (\bibinfo {year} {2010})}\BibitemShut {NoStop}%
\bibitem [{\citenamefont {Kaithakkal}\ \emph {et~al.}(2020)\citenamefont
  {Kaithakkal}, \citenamefont {Kametani},\ and\ \citenamefont
  {Hasegawa}}]{kaithakkal2020dissimilarity}%
  \BibitemOpen
  \bibfield  {author} {\bibinfo {author} {\bibfnamefont {A.~J.}\ \bibnamefont
  {Kaithakkal}}, \bibinfo {author} {\bibfnamefont {Y.}~\bibnamefont
  {Kametani}},\ and\ \bibinfo {author} {\bibfnamefont {Y.}~\bibnamefont
  {Hasegawa}},\ }\bibfield  {title} {\bibinfo {title} {Dissimilarity between
  turbulent heat and momentum transfer induced by a streamwise travelling wave
  of wall blowing and suction},\ }\href@noop {} {\bibfield  {journal} {\bibinfo
   {journal} {Journal of Fluid Mechanics}\ }\textbf {\bibinfo {volume} {886}}
  (\bibinfo {year} {2020})}\BibitemShut {NoStop}%
\bibitem [{\citenamefont {Hammond}\ \emph {et~al.}(1998)\citenamefont
  {Hammond}, \citenamefont {Bewley},\ and\ \citenamefont
  {Moin}}]{hammond1998observed}%
  \BibitemOpen
  \bibfield  {author} {\bibinfo {author} {\bibfnamefont {E.}~\bibnamefont
  {Hammond}}, \bibinfo {author} {\bibfnamefont {T.}~\bibnamefont {Bewley}},\
  and\ \bibinfo {author} {\bibfnamefont {P.}~\bibnamefont {Moin}},\ }\bibfield
  {title} {\bibinfo {title} {Observed mechanisms for turbulence attenuation and
  enhancement in opposition-controlled wall-bounded flows},\ }\href@noop {}
  {\bibfield  {journal} {\bibinfo  {journal} {Physics of Fluids}\ }\textbf
  {\bibinfo {volume} {10}},\ \bibinfo {pages} {2421} (\bibinfo {year}
  {1998})}\BibitemShut {NoStop}%
\bibitem [{\citenamefont {Lee}\ \emph {et~al.}(1998)\citenamefont {Lee},
  \citenamefont {Kim},\ and\ \citenamefont {Choi}}]{lee1998suboptimal}%
  \BibitemOpen
  \bibfield  {author} {\bibinfo {author} {\bibfnamefont {C.}~\bibnamefont
  {Lee}}, \bibinfo {author} {\bibfnamefont {J.}~\bibnamefont {Kim}},\ and\
  \bibinfo {author} {\bibfnamefont {H.}~\bibnamefont {Choi}},\ }\bibfield
  {title} {\bibinfo {title} {Suboptimal control of turbulent channel flow for
  drag reduction},\ }\href@noop {} {\bibfield  {journal} {\bibinfo  {journal}
  {Journal of Fluid Mechanics}\ }\textbf {\bibinfo {volume} {358}},\ \bibinfo
  {pages} {245} (\bibinfo {year} {1998})}\BibitemShut {NoStop}%
\bibitem [{\citenamefont {Han}\ and\ \citenamefont
  {Huang}(2020)}]{han2020active}%
  \BibitemOpen
  \bibfield  {author} {\bibinfo {author} {\bibfnamefont {B.-Z.}\ \bibnamefont
  {Han}}\ and\ \bibinfo {author} {\bibfnamefont {W.-X.}\ \bibnamefont
  {Huang}},\ }\bibfield  {title} {\bibinfo {title} {Active control for drag
  reduction of turbulent channel flow based on convolutional neural networks},\
  }\href@noop {} {\bibfield  {journal} {\bibinfo  {journal} {Physics of
  Fluids}\ }\textbf {\bibinfo {volume} {32}},\ \bibinfo {pages} {095108}
  (\bibinfo {year} {2020})}\BibitemShut {NoStop}%
\bibitem [{\citenamefont {Park}\ and\ \citenamefont
  {Choi}(2020)}]{park2020machine}%
  \BibitemOpen
  \bibfield  {author} {\bibinfo {author} {\bibfnamefont {J.}~\bibnamefont
  {Park}}\ and\ \bibinfo {author} {\bibfnamefont {H.}~\bibnamefont {Choi}},\
  }\bibfield  {title} {\bibinfo {title} {Machine-learning-based feedback
  control for drag reduction in a turbulent channel flow},\ }\href@noop {}
  {\bibfield  {journal} {\bibinfo  {journal} {Journal of Fluid Mechanics}\
  }\textbf {\bibinfo {volume} {904}} (\bibinfo {year} {2020})}\BibitemShut
  {NoStop}%
\bibitem [{\citenamefont {de~Giovanetti}\ \emph {et~al.}(2016)\citenamefont
  {de~Giovanetti}, \citenamefont {Hwang},\ and\ \citenamefont
  {Choi}}]{de2016skin}%
  \BibitemOpen
  \bibfield  {author} {\bibinfo {author} {\bibfnamefont {M.}~\bibnamefont
  {de~Giovanetti}}, \bibinfo {author} {\bibfnamefont {Y.}~\bibnamefont
  {Hwang}},\ and\ \bibinfo {author} {\bibfnamefont {H.}~\bibnamefont {Choi}},\
  }\bibfield  {title} {\bibinfo {title} {Skin-friction generation by attached
  eddies in turbulent channel flow},\ }\href@noop {} {\bibfield  {journal}
  {\bibinfo  {journal} {Journal of Fluid Mechanics}\ }\textbf {\bibinfo
  {volume} {808}},\ \bibinfo {pages} {511} (\bibinfo {year}
  {2016})}\BibitemShut {NoStop}%
\bibitem [{\citenamefont {Wang}\ \emph {et~al.}(2016)\citenamefont {Wang},
  \citenamefont {Huang},\ and\ \citenamefont {Xu}}]{wang2016active}%
  \BibitemOpen
  \bibfield  {author} {\bibinfo {author} {\bibfnamefont {Y.-S.}\ \bibnamefont
  {Wang}}, \bibinfo {author} {\bibfnamefont {W.-X.}\ \bibnamefont {Huang}},\
  and\ \bibinfo {author} {\bibfnamefont {C.-X.}\ \bibnamefont {Xu}},\
  }\bibfield  {title} {\bibinfo {title} {Active control for drag reduction in
  turbulent channel flow: the opposition control schemes revisited},\
  }\href@noop {} {\bibfield  {journal} {\bibinfo  {journal} {Fluid Dynamics
  Research}\ }\textbf {\bibinfo {volume} {48}},\ \bibinfo {pages} {055501}
  (\bibinfo {year} {2016})}\BibitemShut {NoStop}%
\bibitem [{\citenamefont {Samie}\ \emph {et~al.}(2020)\citenamefont {Samie},
  \citenamefont {Baars}, \citenamefont {Rouhi}, \citenamefont {Schlatter},
  \citenamefont {Örlü}, \citenamefont {Marusic},\ and\ \citenamefont
  {Hutchins}}]{SAMIE2020108683}%
  \BibitemOpen
  \bibfield  {author} {\bibinfo {author} {\bibfnamefont {M.}~\bibnamefont
  {Samie}}, \bibinfo {author} {\bibfnamefont {W.}~\bibnamefont {Baars}},
  \bibinfo {author} {\bibfnamefont {A.}~\bibnamefont {Rouhi}}, \bibinfo
  {author} {\bibfnamefont {P.}~\bibnamefont {Schlatter}}, \bibinfo {author}
  {\bibfnamefont {R.}~\bibnamefont {Örlü}}, \bibinfo {author} {\bibfnamefont
  {I.}~\bibnamefont {Marusic}},\ and\ \bibinfo {author} {\bibfnamefont
  {N.}~\bibnamefont {Hutchins}},\ }\bibfield  {title} {\bibinfo {title} {Near
  wall coherence in wall-bounded flows and implications for flow control},\
  }\href
  {https://doi.org/https://doi.org/10.1016/j.ijheatfluidflow.2020.108683}
  {\bibfield  {journal} {\bibinfo  {journal} {International Journal of Heat and
  Fluid Flow}\ }\textbf {\bibinfo {volume} {86}},\ \bibinfo {pages} {108683}
  (\bibinfo {year} {2020})}\BibitemShut {NoStop}%
\bibitem [{\citenamefont {Schoppa}\ and\ \citenamefont
  {Hussain}(1998)}]{schoppa1998large}%
  \BibitemOpen
  \bibfield  {author} {\bibinfo {author} {\bibfnamefont {W.}~\bibnamefont
  {Schoppa}}\ and\ \bibinfo {author} {\bibfnamefont {F.}~\bibnamefont
  {Hussain}},\ }\bibfield  {title} {\bibinfo {title} {A large-scale control
  strategy for drag reduction in turbulent boundary layers},\ }\href@noop {}
  {\bibfield  {journal} {\bibinfo  {journal} {Physics of Fluids}\ }\textbf
  {\bibinfo {volume} {10}},\ \bibinfo {pages} {1049} (\bibinfo {year}
  {1998})}\BibitemShut {NoStop}%
\bibitem [{\citenamefont {Canton}\ \emph {et~al.}(2016)\citenamefont {Canton},
  \citenamefont {{\"O}rl{\"u}}, \citenamefont {Chin},\ and\ \citenamefont
  {Schlatter}}]{canton2016reynolds}%
  \BibitemOpen
  \bibfield  {author} {\bibinfo {author} {\bibfnamefont {J.}~\bibnamefont
  {Canton}}, \bibinfo {author} {\bibfnamefont {R.}~\bibnamefont
  {{\"O}rl{\"u}}}, \bibinfo {author} {\bibfnamefont {C.}~\bibnamefont {Chin}},\
  and\ \bibinfo {author} {\bibfnamefont {P.}~\bibnamefont {Schlatter}},\
  }\bibfield  {title} {\bibinfo {title} {Reynolds number dependence of
  large-scale friction control in turbulent channel flow},\ }\href@noop {}
  {\bibfield  {journal} {\bibinfo  {journal} {Physical Review Fluids}\ }\textbf
  {\bibinfo {volume} {1}},\ \bibinfo {pages} {081501} (\bibinfo {year}
  {2016})}\BibitemShut {NoStop}%
\bibitem [{\citenamefont {Yao}\ \emph {et~al.}(2017)\citenamefont {Yao},
  \citenamefont {Chen}, \citenamefont {Thomas},\ and\ \citenamefont
  {Hussain}}]{Jie2017PRF}%
  \BibitemOpen
  \bibfield  {author} {\bibinfo {author} {\bibfnamefont {J.}~\bibnamefont
  {Yao}}, \bibinfo {author} {\bibfnamefont {X.}~\bibnamefont {Chen}}, \bibinfo
  {author} {\bibfnamefont {F.}~\bibnamefont {Thomas}},\ and\ \bibinfo {author}
  {\bibfnamefont {F.}~\bibnamefont {Hussain}},\ }\bibfield  {title} {\bibinfo
  {title} {Large-scale control strategy for drag reduction in turbulent channel
  flows},\ }\href@noop {} {\bibfield  {journal} {\bibinfo  {journal} {Physical
  Review Fluids}\ }\textbf {\bibinfo {volume} {2}} (\bibinfo {year}
  {2017})}\BibitemShut {NoStop}%
\bibitem [{\citenamefont {Yao}\ \emph {et~al.}(2018)\citenamefont {Yao},
  \citenamefont {Chen},\ and\ \citenamefont {Hussain}}]{yao2018drag}%
  \BibitemOpen
  \bibfield  {author} {\bibinfo {author} {\bibfnamefont {J.}~\bibnamefont
  {Yao}}, \bibinfo {author} {\bibfnamefont {X.}~\bibnamefont {Chen}},\ and\
  \bibinfo {author} {\bibfnamefont {F.}~\bibnamefont {Hussain}},\ }\bibfield
  {title} {\bibinfo {title} {Drag control in wall-bounded turbulent flows via
  spanwise opposed wall-jet forcing},\ }\href@noop {} {\bibfield  {journal}
  {\bibinfo  {journal} {Journal of Fluid Mechanics}\ }\textbf {\bibinfo
  {volume} {852}},\ \bibinfo {pages} {678} (\bibinfo {year}
  {2018})}\BibitemShut {NoStop}%
\bibitem [{\citenamefont {Iuso}\ \emph {et~al.}(2002)\citenamefont {Iuso},
  \citenamefont {Onorato}, \citenamefont {Spazzini},\ and\ \citenamefont
  {Di~Cicca}}]{iuso2002wall}%
  \BibitemOpen
  \bibfield  {author} {\bibinfo {author} {\bibfnamefont {G.}~\bibnamefont
  {Iuso}}, \bibinfo {author} {\bibfnamefont {M.}~\bibnamefont {Onorato}},
  \bibinfo {author} {\bibfnamefont {P.~G.}\ \bibnamefont {Spazzini}},\ and\
  \bibinfo {author} {\bibfnamefont {G.~M.}\ \bibnamefont {Di~Cicca}},\
  }\bibfield  {title} {\bibinfo {title} {Wall turbulence manipulation by
  large-scale streamwise vortices},\ }\href@noop {} {\bibfield  {journal}
  {\bibinfo  {journal} {Journal of Fluid Mechanics}\ }\textbf {\bibinfo
  {volume} {473}},\ \bibinfo {pages} {23} (\bibinfo {year} {2002})}\BibitemShut
  {NoStop}%
\bibitem [{\citenamefont {Cannata}\ \emph {et~al.}(2020)\citenamefont
  {Cannata}, \citenamefont {Cafiero},\ and\ \citenamefont
  {Iuso}}]{cannata2020large}%
  \BibitemOpen
  \bibfield  {author} {\bibinfo {author} {\bibfnamefont {M.}~\bibnamefont
  {Cannata}}, \bibinfo {author} {\bibfnamefont {G.}~\bibnamefont {Cafiero}},\
  and\ \bibinfo {author} {\bibfnamefont {G.}~\bibnamefont {Iuso}},\ }\bibfield
  {title} {\bibinfo {title} {Large-scale forcing of a turbulent channel flow
  through spanwise synthetic jets},\ }\href@noop {} {\bibfield  {journal}
  {\bibinfo  {journal} {AIAA Journal}\ }\textbf {\bibinfo {volume} {58}},\
  \bibinfo {pages} {2042} (\bibinfo {year} {2020})}\BibitemShut {NoStop}%
\bibitem [{\citenamefont {Wong}\ \emph {et~al.}(2015)\citenamefont {Wong},
  \citenamefont {Zhou}, \citenamefont {Li},\ and\ \citenamefont
  {Li}}]{wong2015active}%
  \BibitemOpen
  \bibfield  {author} {\bibinfo {author} {\bibfnamefont {C.~W.}\ \bibnamefont
  {Wong}}, \bibinfo {author} {\bibfnamefont {Y.}~\bibnamefont {Zhou}}, \bibinfo
  {author} {\bibfnamefont {Y.}~\bibnamefont {Li}},\ and\ \bibinfo {author}
  {\bibfnamefont {Y.}~\bibnamefont {Li}},\ }\bibfield  {title} {\bibinfo
  {title} {Active drag reduction in a turbulent boundary layer based on
  plasma-actuator-generated streamwise vortices},\ }in\ \href@noop {} {\emph
  {\bibinfo {booktitle} {Proceeding of the 9th International Symposium on
  Turbulence and Shear Flow Phenomena}}}\ (\bibinfo {year} {2015})\BibitemShut
  {NoStop}%
\bibitem [{\citenamefont {Corke}\ and\ \citenamefont
  {Thomas}(2018)}]{corke2018active}%
  \BibitemOpen
  \bibfield  {author} {\bibinfo {author} {\bibfnamefont {T.~C.}\ \bibnamefont
  {Corke}}\ and\ \bibinfo {author} {\bibfnamefont {F.~O.}\ \bibnamefont
  {Thomas}},\ }\bibfield  {title} {\bibinfo {title} {Active and passive
  turbulent boundary-layer drag reduction},\ }\href@noop {} {\bibfield
  {journal} {\bibinfo  {journal} {AIAA Journal}\ }\textbf {\bibinfo {volume}
  {56}},\ \bibinfo {pages} {3835} (\bibinfo {year} {2018})}\BibitemShut
  {NoStop}%
\bibitem [{\citenamefont {Abbassi}\ \emph {et~al.}(2017)\citenamefont
  {Abbassi}, \citenamefont {Baars}, \citenamefont {Hutchins},\ and\
  \citenamefont {Marusic}}]{abbassi2017skin}%
  \BibitemOpen
  \bibfield  {author} {\bibinfo {author} {\bibfnamefont {M.}~\bibnamefont
  {Abbassi}}, \bibinfo {author} {\bibfnamefont {W.}~\bibnamefont {Baars}},
  \bibinfo {author} {\bibfnamefont {N.}~\bibnamefont {Hutchins}},\ and\
  \bibinfo {author} {\bibfnamefont {I.}~\bibnamefont {Marusic}},\ }\bibfield
  {title} {\bibinfo {title} {Skin-friction drag reduction in a
  high-reynolds-number turbulent boundary layer via real-time control of
  large-scale structures},\ }\href@noop {} {\bibfield  {journal} {\bibinfo
  {journal} {International Journal of Heat and Fluid Flow}\ }\textbf {\bibinfo
  {volume} {67}},\ \bibinfo {pages} {30} (\bibinfo {year} {2017})}\BibitemShut
  {NoStop}%
\bibitem [{\citenamefont {Hwang}\ \emph {et~al.}(2016)\citenamefont {Hwang},
  \citenamefont {Lee}, \citenamefont {Sung},\ and\ \citenamefont
  {Zaki}}]{hwang2016inner}%
  \BibitemOpen
  \bibfield  {author} {\bibinfo {author} {\bibfnamefont {J.}~\bibnamefont
  {Hwang}}, \bibinfo {author} {\bibfnamefont {J.}~\bibnamefont {Lee}}, \bibinfo
  {author} {\bibfnamefont {H.~J.}\ \bibnamefont {Sung}},\ and\ \bibinfo
  {author} {\bibfnamefont {T.~A.}\ \bibnamefont {Zaki}},\ }\bibfield  {title}
  {\bibinfo {title} {Inner--outer interactions of large-scale structures in
  turbulent channel flow},\ }\href@noop {} {\bibfield  {journal} {\bibinfo
  {journal} {Journal of Fluid Mechanics}\ }\textbf {\bibinfo {volume} {790}},\
  \bibinfo {pages} {128} (\bibinfo {year} {2016})}\BibitemShut {NoStop}%
\bibitem [{\citenamefont {Chen}\ \emph {et~al.}(2021)\citenamefont {Chen},
  \citenamefont {Yao},\ and\ \citenamefont {Hussain}}]{chen2021}%
  \BibitemOpen
  \bibfield  {author} {\bibinfo {author} {\bibfnamefont {X.}~\bibnamefont
  {Chen}}, \bibinfo {author} {\bibfnamefont {J.}~\bibnamefont {Yao}},\ and\
  \bibinfo {author} {\bibfnamefont {F.}~\bibnamefont {Hussain}},\ }\bibfield
  {title} {\bibinfo {title} {Theoretical framework for energy flux analysis of
  channels under drag control},\ }\href
  {https://doi.org/10.1103/PhysRevFluids.6.013902} {\bibfield  {journal}
  {\bibinfo  {journal} {Phys. Rev. Fluids}\ }\textbf {\bibinfo {volume} {6}},\
  \bibinfo {pages} {013902} (\bibinfo {year} {2021})}\BibitemShut {NoStop}%
\bibitem [{\citenamefont {Lee}\ and\ \citenamefont
  {Moser}(2015)}]{lee2015direct}%
  \BibitemOpen
  \bibfield  {author} {\bibinfo {author} {\bibfnamefont {M.}~\bibnamefont
  {Lee}}\ and\ \bibinfo {author} {\bibfnamefont {R.~D.}\ \bibnamefont
  {Moser}},\ }\bibfield  {title} {\bibinfo {title} {Direct numerical simulation
  of turbulent channel flow up to re 5200},\ }\href@noop {} {\bibfield
  {journal} {\bibinfo  {journal} {Journal of Fluid Mechanics}\ }\textbf
  {\bibinfo {volume} {774}},\ \bibinfo {pages} {395} (\bibinfo {year}
  {2015})}\BibitemShut {NoStop}%
\bibitem [{\citenamefont {Kim}\ \emph {et~al.}(1987)\citenamefont {Kim},
  \citenamefont {Moin},\ and\ \citenamefont {Moser}}]{kim1987turbulence}%
  \BibitemOpen
  \bibfield  {author} {\bibinfo {author} {\bibfnamefont {J.}~\bibnamefont
  {Kim}}, \bibinfo {author} {\bibfnamefont {P.}~\bibnamefont {Moin}},\ and\
  \bibinfo {author} {\bibfnamefont {R.}~\bibnamefont {Moser}},\ }\bibfield
  {title} {\bibinfo {title} {Turbulence statistics in fully developed channel
  flow at low reynolds number},\ }\href@noop {} {\bibfield  {journal} {\bibinfo
   {journal} {Journal of fluid mechanics}\ }\textbf {\bibinfo {volume} {177}},\
  \bibinfo {pages} {133} (\bibinfo {year} {1987})}\BibitemShut {NoStop}%
\bibitem [{\citenamefont {Reynolds}\ and\ \citenamefont
  {Hussain}(1972)}]{reynolds1972mechanics}%
  \BibitemOpen
  \bibfield  {author} {\bibinfo {author} {\bibfnamefont {W.}~\bibnamefont
  {Reynolds}}\ and\ \bibinfo {author} {\bibfnamefont {A.}~\bibnamefont
  {Hussain}},\ }\bibfield  {title} {\bibinfo {title} {The mechanics of an
  organized wave in turbulent shear flow. part 3. theoretical models and
  comparisons with experiments},\ }\href@noop {} {\bibfield  {journal}
  {\bibinfo  {journal} {Journal of Fluid Mechanics}\ }\textbf {\bibinfo
  {volume} {54}},\ \bibinfo {pages} {263} (\bibinfo {year} {1972})}\BibitemShut
  {NoStop}%
\bibitem [{\citenamefont {Stroh}\ \emph {et~al.}(2015)\citenamefont {Stroh},
  \citenamefont {Frohnapfel}, \citenamefont {Schlatter},\ and\ \citenamefont
  {Hasegawa}}]{stroh2015comparison}%
  \BibitemOpen
  \bibfield  {author} {\bibinfo {author} {\bibfnamefont {A.}~\bibnamefont
  {Stroh}}, \bibinfo {author} {\bibfnamefont {B.}~\bibnamefont {Frohnapfel}},
  \bibinfo {author} {\bibfnamefont {P.}~\bibnamefont {Schlatter}},\ and\
  \bibinfo {author} {\bibfnamefont {Y.}~\bibnamefont {Hasegawa}},\ }\bibfield
  {title} {\bibinfo {title} {A comparison of opposition control in turbulent
  boundary layer and turbulent channel flow},\ }\href@noop {} {\bibfield
  {journal} {\bibinfo  {journal} {Physics of fluids}\ }\textbf {\bibinfo
  {volume} {27}},\ \bibinfo {pages} {075101} (\bibinfo {year}
  {2015})}\BibitemShut {NoStop}%
\bibitem [{\citenamefont {Deng}\ \emph {et~al.}(2014)\citenamefont {Deng},
  \citenamefont {Xu}, \citenamefont {Huang},\ and\ \citenamefont
  {Cui}}]{deng2014strengthened}%
  \BibitemOpen
  \bibfield  {author} {\bibinfo {author} {\bibfnamefont {B.-Q.}\ \bibnamefont
  {Deng}}, \bibinfo {author} {\bibfnamefont {C.-X.}\ \bibnamefont {Xu}},
  \bibinfo {author} {\bibfnamefont {W.-X.}\ \bibnamefont {Huang}},\ and\
  \bibinfo {author} {\bibfnamefont {G.-X.}\ \bibnamefont {Cui}},\ }\bibfield
  {title} {\bibinfo {title} {Strengthened opposition control for skin-friction
  reduction in wall-bounded turbulent flows},\ }\href@noop {} {\bibfield
  {journal} {\bibinfo  {journal} {Journal of Turbulence}\ }\textbf {\bibinfo
  {volume} {15}},\ \bibinfo {pages} {122} (\bibinfo {year} {2014})}\BibitemShut
  {NoStop}%
\bibitem [{\citenamefont {Jeong}\ and\ \citenamefont
  {Hussain}(1995)}]{jeong1995identification}%
  \BibitemOpen
  \bibfield  {author} {\bibinfo {author} {\bibfnamefont {J.}~\bibnamefont
  {Jeong}}\ and\ \bibinfo {author} {\bibfnamefont {F.}~\bibnamefont
  {Hussain}},\ }\bibfield  {title} {\bibinfo {title} {On the identification of
  a vortex},\ }\href@noop {} {\bibfield  {journal} {\bibinfo  {journal}
  {Journal of fluid mechanics}\ }\textbf {\bibinfo {volume} {285}},\ \bibinfo
  {pages} {69} (\bibinfo {year} {1995})}\BibitemShut {NoStop}%
\bibitem [{\citenamefont {Fukagata}\ \emph {et~al.}(2002)\citenamefont
  {Fukagata}, \citenamefont {Iwamoto},\ and\ \citenamefont
  {Kasagi}}]{fukagata2002contribution}%
  \BibitemOpen
  \bibfield  {author} {\bibinfo {author} {\bibfnamefont {K.}~\bibnamefont
  {Fukagata}}, \bibinfo {author} {\bibfnamefont {K.}~\bibnamefont {Iwamoto}},\
  and\ \bibinfo {author} {\bibfnamefont {N.}~\bibnamefont {Kasagi}},\
  }\bibfield  {title} {\bibinfo {title} {Contribution of reynolds stress
  distribution to the skin friction in wall-bounded flows},\ }\href@noop {}
  {\bibfield  {journal} {\bibinfo  {journal} {Physics of Fluids
  (1994-present)}\ }\textbf {\bibinfo {volume} {14}},\ \bibinfo {pages} {L73}
  (\bibinfo {year} {2002})}\BibitemShut {NoStop}%
\bibitem [{\citenamefont {Rastegari}\ and\ \citenamefont
  {Akhavan}(2015)}]{rastegari2015mechanism}%
  \BibitemOpen
  \bibfield  {author} {\bibinfo {author} {\bibfnamefont {A.}~\bibnamefont
  {Rastegari}}\ and\ \bibinfo {author} {\bibfnamefont {R.}~\bibnamefont
  {Akhavan}},\ }\bibfield  {title} {\bibinfo {title} {On the mechanism of
  turbulent drag reduction with super-hydrophobic surfaces},\ }\href@noop {}
  {\bibfield  {journal} {\bibinfo  {journal} {Journal of Fluid Mechanics}\
  }\textbf {\bibinfo {volume} {773}} (\bibinfo {year} {2015})}\BibitemShut
  {NoStop}%
\bibitem [{\citenamefont {Renard}\ and\ \citenamefont
  {Deck}(2016)}]{renard2016theoretical}%
  \BibitemOpen
  \bibfield  {author} {\bibinfo {author} {\bibfnamefont {N.}~\bibnamefont
  {Renard}}\ and\ \bibinfo {author} {\bibfnamefont {S.}~\bibnamefont {Deck}},\
  }\bibfield  {title} {\bibinfo {title} {A theoretical decomposition of mean
  skin friction generation into physical phenomena across the boundary layer},\
  }\href@noop {} {\bibfield  {journal} {\bibinfo  {journal} {Journal of Fluid
  Mechanics}\ }\textbf {\bibinfo {volume} {790}},\ \bibinfo {pages} {339}
  (\bibinfo {year} {2016})}\BibitemShut {NoStop}%
\bibitem [{\citenamefont {Ricco}\ \emph {et~al.}(2012)\citenamefont {Ricco},
  \citenamefont {Ottonelli}, \citenamefont {Hasegawa},\ and\ \citenamefont
  {Quadrio}}]{ricco2012changes}%
  \BibitemOpen
  \bibfield  {author} {\bibinfo {author} {\bibfnamefont {P.}~\bibnamefont
  {Ricco}}, \bibinfo {author} {\bibfnamefont {C.}~\bibnamefont {Ottonelli}},
  \bibinfo {author} {\bibfnamefont {Y.}~\bibnamefont {Hasegawa}},\ and\
  \bibinfo {author} {\bibfnamefont {M.}~\bibnamefont {Quadrio}},\ }\bibfield
  {title} {\bibinfo {title} {Changes in turbulent dissipation in a channel flow
  with oscillating walls},\ }\href@noop {} {\bibfield  {journal} {\bibinfo
  {journal} {Journal of Fluid Mechanics}\ }\textbf {\bibinfo {volume} {700}},\
  \bibinfo {pages} {77} (\bibinfo {year} {2012})}\BibitemShut {NoStop}%
\bibitem [{\citenamefont {Gatti}\ \emph {et~al.}(2018)\citenamefont {Gatti},
  \citenamefont {Cimarelli}, \citenamefont {Hasegawa}, \citenamefont
  {Frohnapfel},\ and\ \citenamefont {Quadrio}}]{gatti2018global}%
  \BibitemOpen
  \bibfield  {author} {\bibinfo {author} {\bibfnamefont {D.}~\bibnamefont
  {Gatti}}, \bibinfo {author} {\bibfnamefont {A.}~\bibnamefont {Cimarelli}},
  \bibinfo {author} {\bibfnamefont {Y.}~\bibnamefont {Hasegawa}}, \bibinfo
  {author} {\bibfnamefont {B.}~\bibnamefont {Frohnapfel}},\ and\ \bibinfo
  {author} {\bibfnamefont {M.}~\bibnamefont {Quadrio}},\ }\bibfield  {title}
  {\bibinfo {title} {Global energy fluxes in turbulent channels with flow
  control},\ }\href@noop {} {\bibfield  {journal} {\bibinfo  {journal} {Journal
  of Fluid Mechanics}\ }\textbf {\bibinfo {volume} {857}},\ \bibinfo {pages}
  {345} (\bibinfo {year} {2018})}\BibitemShut {NoStop}%
\bibitem [{\citenamefont {Alessio}\ \emph {et~al.}(2021)\citenamefont
  {Alessio}, \citenamefont {Francesco},\ and\ \citenamefont
  {Alfredo}}]{zonta2021}%
  \BibitemOpen
  \bibfield  {author} {\bibinfo {author} {\bibfnamefont {R.}~\bibnamefont
  {Alessio}}, \bibinfo {author} {\bibfnamefont {Z.}~\bibnamefont {Francesco}},\
  and\ \bibinfo {author} {\bibfnamefont {S.}~\bibnamefont {Alfredo}},\
  }\bibfield  {title} {\bibinfo {title} {Energy balance in lubricated
  drag-reduced turbulent channel flow},\ }\href@noop {} {\bibfield  {journal}
  {\bibinfo  {journal} {Journal of Fluid Mechanics}\ }\textbf {\bibinfo
  {volume} {Accepted}} (\bibinfo {year} {2021})}\BibitemShut {NoStop}%
\bibitem [{\citenamefont {Marusic}\ \emph {et~al.}(2010)\citenamefont
  {Marusic}, \citenamefont {Mathis},\ and\ \citenamefont
  {Hutchins}}]{marusic2010predictive}%
  \BibitemOpen
  \bibfield  {author} {\bibinfo {author} {\bibfnamefont {I.}~\bibnamefont
  {Marusic}}, \bibinfo {author} {\bibfnamefont {R.}~\bibnamefont {Mathis}},\
  and\ \bibinfo {author} {\bibfnamefont {N.}~\bibnamefont {Hutchins}},\
  }\bibfield  {title} {\bibinfo {title} {Predictive model for wall-bounded
  turbulent flow},\ }\href@noop {} {\bibfield  {journal} {\bibinfo  {journal}
  {Science}\ }\textbf {\bibinfo {volume} {329}},\ \bibinfo {pages} {193}
  (\bibinfo {year} {2010})}\BibitemShut {NoStop}%
\end{thebibliography}%

\end{document}